\documentclass[preprint,showpacs,preprintnumbers,amsmath,amssymb]{revtex4}
\usepackage{graphicx}
\usepackage{dcolumn}
\usepackage{bm}

\begin{document}


\title{Stability of Bose-Einstein condensates
\\ in a Kronig-Penney potential}

\author{Ippei Danshita$^{1,2}$}
\email{danshita@nist.gov}
\author{Shunji Tsuchiya$^3$}
\affiliation{$^1$National Institute of Standards and Technology, Gaithersburg, MD 20899, USA
\\
$^2$Department of Physics, Waseda University, \=Okubo, Shinjuku, Tokyo, 169-8555, Japan
\\
$^3$Dipartimento di Fisica, Universit\'a di Trento and CNR-INFM BEC Center, I-38050 Povo, Trento, Italy
}
\date{\today}

\begin{abstract}
We study the stability of Bose-Einstein condensates with superfluid currents in a one-dimensional periodic potential.
By using the Kronig-Penney model, the condensate and Bogoliubov bands are analytically calculated and the stability of condensates in a periodic potential is discussed.
The Landau and dynamical instabilities occur in a Kronig-Penney potential when the quasimomentum of the condensate exceeds certain critical values as in a sinusoidal potential.
It is found that the onsets of the Landau and dynamical
instabilities coincide with the point where the perfect transmission
of low energy excitations through each potential barrier is forbidden.
The Landau instability is caused by the excitations with small $q$ and
the dynamical instability is caused by the excitations with $q=\pi/a$ at their onsets, 
where $q$ is the quasimomentum of excitation and $a$ is the lattice constant.
A swallow-tail energy loop appears at the edge of the first condensate band when the mean-field energy is sufficiently larger than the strength of the periodic potential. 
We find that the upper portion of the swallow-tail is always dynamically unstable, but the second Bogoliubov band has a phonon spectrum reflecting the positive effective mass.
\end{abstract}

\pacs{03.75.Lm \ 05.30.Jn \ 03.75.Kk}
\keywords{Bose-Einstein condensation, Bogoliubov equations, optical lattice, Kronig-Penney potential, elementary excitation, dynamical instability}
\maketitle

\section{\label{sec:level1}Introduction}
Since its first achievement~\cite{rf:origin}, Bose-Einstein condensates (BECs) in an optical lattice have been studied vigorously~\cite{rf:rev}. Various interesting phenomena have been observed in such systems, including the superfluid-Mott insulator transition~\cite{rf:SI,rf:SINa}, formation of bright gap solitons~\cite{rf:gapsol}, the Josephson effect~\cite{rf:jose1,rf:jose2}, and breakdown of superfluidity~\cite{rf:inst1,rf:inst2,rf:inst3,rf:moving1,rf:moving2}. 
In particular, breakdown of superfluidity has attracted much attention.
Recently, the stability of BECs in an optical lattice moving at constant velocity has been experimentally investigated and it was demonstrated that the Landau and dynamical instabilities occur when the velocity of the lattice potential exceeds certain critical values~\cite{rf:moving2}.

Swallow-tail energy loops have been theoretically found in the band structure of BECs in a periodic potential~\cite{rf:loop,rf:swallow,rf:pesmith,rf:mueller,rf:seaman1,rf:seaman2} and attracted much interests. 
Mueller pointed out that a Bloch state in the upper portion of swallow-tails corresponds to a saddle point in the energy landscape and it is dynamically unstable~\cite{rf:mueller}.
In fact, Seaman {\it et al.} have shown the instability of a Bloch state in the upper portion of swallow-tails by numerically solving the time-dependent Gross-Pitaevskii (GP) equation~\cite{rf:seaman1}.
However, the stability of swallow-tails has not been fully investigated because a detail study of linear stability problem is still lacking. One needs to identify the unstable modes of excitations by solving the Bogoliubov equations.


In this paper, in order to understand the superfluidity of BECs in an optical lattice and swallow-tail energy loops, we study BECs in a periodic potential by using the Kronig-Penney (KP) model. It is well-known in solid state physics that the KP model is useful for understanding the band structure of electrons~\cite{rf:kittel}. The KP model is also useful for the study of BECs in a periodic potential and has remarkable advantages. 
First, it allows one to calculate the whole band structure of condensate energy and excitation spectrum analytically. Another advantage is that the band structure of excitation spectrum can be calculated from the tunneling problem of Bogoliubov excitations through a single potential barrier. The tunneling problem of Bogoliubov excitations has been studied in several papers and it was found that low energy Bogoliubov excitations exhibit {\it anomalous tunneling} i.e. a potential barrier is transparent for them~\cite{rf:kov,rf:antun}. 
In our previous work, the excitation spectrum of current-free condensates has been calculated using the KP model, and it has been shown that the anomalous tunneling is crucial to the phonon spectrum of excitations~\cite{rf:wareware}.  



In the present paper, we extend the previous work to calculate the excitation spectrum of condensates with supercurrents. We solve the Bogoliubov equations using the solution of the tunneling problem of Bogoliubov excitations~\cite{rf:ware}, and discuss the stability of condensates with supercurrents.
We will show that the Landau and dynamical instabilities occur when the quasimomentum of the condensate exceeds certain critical values. We identify the unstable modes causing the Landau and dynamical instabilities.
We will also show that the Bloch state of the upper portion of swallow-tail is dynamically unstable, but the excitation spectrum has a phonon branch due to the positive effective mass.



The outline of the paper is as follows.
In Sec. \!II, we introduce our model and formulation based on the Gross-Pitaevskii mean-field theory.
In Sec. \!III, we solve the time-independent GP equation and calculate the first condensate band, group velocity and effective mass. The condition for the presence of swallow-tails are discussed.
In Sec. \!IV, we solve the Bogoliubov equations and calculate the band structure of the excitation spectrum.
In Sec. \!V, we discuss the stability of BECs with superfluid currents in a periodic potential.
In Sec. \!VI, we summarize our results. 
Some details of our calculation in Sec. \!III and \!IV are given in Appendix.
\section{model and formulation}
We consider a BEC at the absolute zero of temperature confined in a combined potential of a box-shaped trap and a one-dimensional periodic potential.
A box-shaped trap was realized in a recent experiment with a radial harmonic confinement and end caps~\cite{rf:box}.
The frequency of the radial harmonic potential is assumed to be much larger than the excitation energy in the axial direction so that the one-dimensional treatment of the problem is justified~\cite{rf:wareware}.
It is also assumed that the axial size of the system is so large that the effect of the edge of the system can be neglected.
We consider condensates with supercurrents flowing through the periodic potential. 
This situation can be realized by an optical lattice consisting of two counter-propagating laser beams with a frequency difference and moving in a constant velocity~\cite{rf:moving1,rf:moving2}.

\begin{figure}[tb]
\includegraphics[width=7.5cm,height=3.5cm]{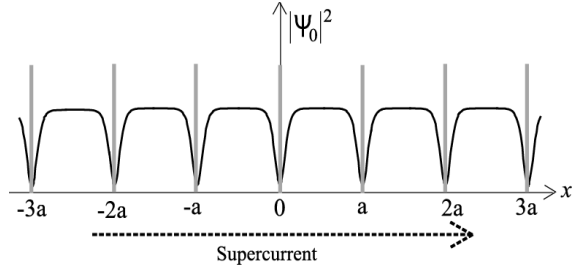}
\caption{\label{fig:kpcon}
Schematic picture of a current-carrying condensate in a KP potential.
The solid lines, gray bars and dashed arrow express the condensate density $|\Psi_0(x)|^2$, potential barriers and supercurrent, respectively.
}
\end{figure}

The periodic potential is assumed to be an array of $\delta$-function potential barriers,
  \begin{eqnarray}
  V(x)=V_0\sum_{j}\delta(x-ja), \label{KPpotential}
  \end{eqnarray}
where $a$ is the lattice constant, and $V_0$ is the potential strength. Note that $V_0$ and $a$ are independent parameters.
Equation (\ref{KPpotential}) is a special type of the KP potential.
A schematic picture of the system is shown in Fig. \ref{fig:kpcon}.
Although the periodic potential of optical lattices in experiments is sinusoidal, the KP model provides us with valuable insights. BECs in a KP potential have the same properties as those in a sinusoidal potential, such as the band structure~\cite{rf:seaman1,rf:seaman2} and the excitation spectrum of a current-free condensate~\cite{rf:wareware}. As we discuss later, the Landau and dynamical instabilities of BECs occur in a KP potential as in a sinusoidal potential.

Our formulation of the problem is based on the Gross-Pitaevskii mean-field theory~\cite{rf:BEC,rf:mark}.
The condensate wave function $\Psi(x,t)$ obeys the time-dependent GP equation,
  \begin{eqnarray}
  {\rm i}\hbar\frac{\partial\Psi}{\partial t}=
  \left[-\frac{\hbar^2}{2m}\frac{d^2}{d x^2}+V(x)+g|\Psi|^2
  \right]\Psi,
  \label{eq:tdGPE}
  \end{eqnarray}
where $m$ is the mass of the atoms and $g$ is the mean-field coupling constant. 
The coupling constant in 1D is given by $g=\frac{2\hbar^2a_0}{ma_{\perp}^2}$~\cite{rf:oned}, where $a_0$ is the $s$-wave scattering length and $a_{\perp}$ is the harmonic oscillator length of the radial confinement.
The static solution $\Psi_0(x)$ of Eq. (\ref{eq:tdGPE}) satisfies the time-independent GP equation
  \begin{eqnarray}
  \left[-\frac{\hbar^2}{2m}\frac{d^2}{d x^2}-\mu+V(x)
  +g|\Psi_0(x)|^2\right]\Psi_0(x) =0,\label{eq:sGPE}
  \end{eqnarray}
where $\mu$ is the chemical potential.
The normalization condition is given by
\begin{eqnarray}
  \int_{ja}^{(j+1)a}\left|\Psi_0(x)\right|^2dx=N_{\rm C}.\label{eq:normal}
  \end{eqnarray}
$N_{\rm C}$ is the number of condensate atoms per site.
The energy of the condensate per site is given by
  \begin{eqnarray}
  E = \int_{ja}^{(j+1)a}dx\Psi_0^{\ast}
  \left[-\frac{\hbar^2}{2m}\frac{d^2}{dx^2}+V(x)+\frac{g}{2}|\Psi_0|^2
  \right]\Psi_0.\label{eq:mean_ener}
  \end{eqnarray}
Considering small fluctuations from $\Psi_0(x)$, the condensate wave function can be written as
  \begin{eqnarray}
  \Psi(x,t)={\rm e}^{-\frac{{\rm i}\mu t}{\hbar}}
  \left[\Psi_0(x)
  +u(x){\rm e}^{-\frac{{\rm i}\varepsilon t}{\hbar}}
  -v(x)^{\ast}{\rm e}^{\frac{{\rm i}\varepsilon t}{\hbar}}\right].\label{eq:fluctuation}
  \end{eqnarray}
Substituting Eq. \!(\ref{eq:fluctuation}) into Eq. \!(\ref{eq:tdGPE}) and keeping only linear terms of the fluctuation, $\left(u(x),v(x)\right)$ obeys the Bogoliubov equations.
       \begin{eqnarray}
               \left(
                 \begin{array}{cc}
                 H_0 & -g{\Psi_0(x)}^2 \\
                 g{\Psi_0(x)}^{\ast2} & -H_0
                 \end{array}
               \right) 
               \left(
                 \begin{array}{cc}
                 u(x) \\ v(x)
                 \end{array}
               \right)
               = \varepsilon\left(
                 \begin{array}{cc}
                   u(x) \\ v(x)
                 \end{array}
               \right),  \label{eq:BdGE}\\
               H_0 \equiv -\frac{\hbar^2}{2m}\frac{d^2}{d x^2}
               -\mu+V\left(x\right)+2g|\Psi_0|^2,
       \end{eqnarray}
$\varepsilon/\hbar$ and $\left(u(x),v(x)\right)$ describe the frequency and amplitude of the collective mode of the condensate.

Asserting the normalization condition
  \begin{eqnarray}
  \int dx\left(|u(x)|^2-|v(x)|^2\right)=1,\label{eq:normal_ex}
  \end{eqnarray}
one can quantize the collective modes and regard them as elementary excitations~\cite{rf:BEC,rf:mark}.
We hereafter call $\varepsilon$ and $\left(u(x),v(x)\right)$ as the energy and the wave function of the elementary excitation.
Strictly speaking, {\it elementary excitations with complex energies} do not exist, because the normalization condition Eq. (\ref{eq:normal_ex}) is not satisfied if $\varepsilon$ is not real~\cite{rf:wuniu2,rf:iigaya,rf:BEC}.
However, we use this term by implicitly meaning {\it collective modes with complex frequencies}.

The stability of BECs can be studied by calculating the excitation energy.
The appearance of excitations with negative energies signals the Landau instability.
It reveals that the solution $\Psi_0$ of the time-independent GP equation does not correspond to a local minimum in the energy landscape~\cite{rf:wuniu1,rf:wuniu2,rf:BEC}.
This means that the process of spontaneous creation of excitations can take place and the system becomes unstable~\cite{rf:landau}.
On the other hand, the appearance of excitations with complex energies signals the dynamical instability which means exponential growth of the fluctuation in time.
The detail of the difference between the Landau and dynamical instabilities has been discussed in Refs.~\cite{rf:iigaya,rf:konabe1}.
\section{Condensate band and swallow-tail}
In this section, we shall obtain the Bloch state solution of Eq. \!(\ref{eq:sGPE}) and calculate the energy of the condensate $E$, group velocity $v_{\rm g}$ and effective mass $m^{\ast}$ as functions of the condensate quasimomentum $K$. Assuming that the lattice constant is much larger than the healing length, analytic expressions of these quantities are obtained. Hereafter, we call the band structure of the energy of the condensate $E(K)$ as the {\it condensate band}.

\subsection{Condensate wave function}
The condensate wave function can be written as $\Psi_0(x)=\sqrt{n_0}A(x){\rm e}^{{\rm i}S(x)}$, where $A(x)$ and $S(x)$ mean the amplitude and phase of the condensate. $A(x)$ is normalized by the density at the center of each site $n_0\equiv |\Psi_0\left(\left(j+1/2\right)a\right)|^2$.
Thus, Eq. \!(\ref{eq:sGPE}) is reduced to a set of equations as (we set $\hbar=1$ from now on)
   \begin{eqnarray}
      -\frac{1}{2m}\frac{d^2A}{dx^2}+\frac{Q^2}{2m}A^{-3}
      +V(x)A-\mu A+gn_0 A^3\!=\!0,      \label{eq:ampl}
   \end{eqnarray}
   \begin{eqnarray}
      A^2\frac{dS}{dx}=Q.\label{eq:conti}
   \end{eqnarray}
Equation \!(\ref{eq:conti}) is the equation of continuity. $Q$ describes the superfluid momentum and the supercurrent is given by $\frac{n_0 Q}{m}$. They are conserved for the static solution and independent of $x$. Note that the effect of the $\delta$-function potential barriers appears only through the boundary condition at $x=ja$.

\begin{figure}[tb]
\includegraphics[width=7.5cm,height=5.0cm]{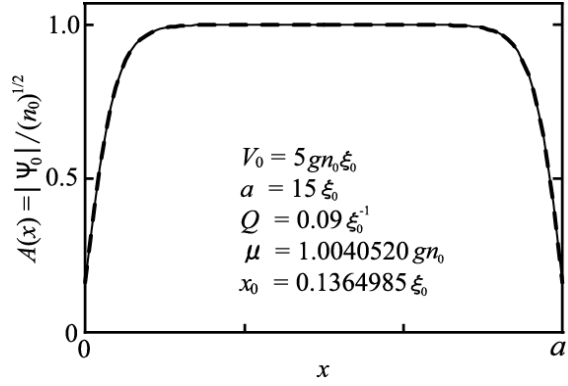}
\caption{\label{fig:ampl}
Amplitude $A(x)$ of the condensate wave function when $V_0=5gn_0\xi_0$, $a=15\xi_0$, and $Q=0.09\xi_0^{-1}$.
The dashed and solid lines represent the exact solution Eq. \!(\ref{eq:dens}) and the approximate solution Eq. \!(\ref{eq:dens_th}), respectively.
}
\end{figure}
\begin{figure}[tb]
\includegraphics[width=7.5cm,height=5.0cm]{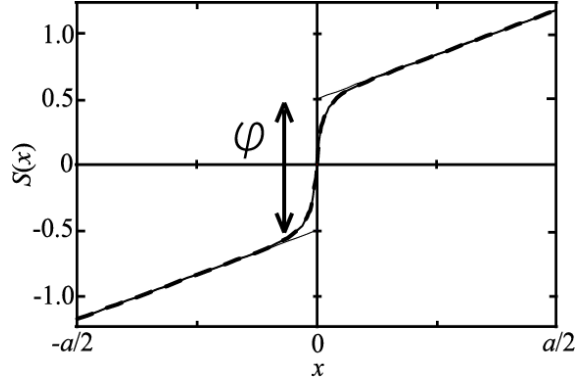}
\caption{\label{fig:phase}
Phase $S(x)$ of the condensate wave function when $V_0=5gn_0\xi_0$, $a=15\xi_0$, and $Q=0.09\xi_0^{-1}$.
The dashed line, thick solid line and thin solid line represent the exact solution Eq. \!(\ref{eq:phasedef}), approximate solution Eq. \!(\ref{eq:phase_th}), and asymptotic solution $Qx+{\rm sgn}(x)\varphi/2$, respectively.
}
\end{figure}
Throughout the present paper, we assume that the condensate sits in the first condensate band. In this case, the condensate wave function has no node in a single well \cite{rf:seaman1} as shown in Fig. 2. 

Since the amplitude of the condensate wave function in the first band has a maximum at the center of each well, $A(x)$ satisfies the boundary condition
  \begin{eqnarray}
  \left.\frac{dA}{dx}\right|_{(j+1/2)a}=0.
  \label{eq:centerd}
  \end{eqnarray}
From the definition of the central density $n_0$, $A(x)$ also satisfies
  \begin{eqnarray}
  A\left(\left(j+\frac{1}{2}\right)a\right) 
  = 1.\label{eq:center}
  \end{eqnarray}
The boundary condition at $x=ja$ can be obtained by integrating Eq. \!(\ref{eq:ampl}) from $ja-0$ to $ja+0$ as
  \begin{eqnarray}
  \left.\frac{dA^2}{dx}\right|_{ja+0}=
  \left.\frac{dA^2}{dx}\right|_{ja-0}+4mV_0 A^2(ja).
  \label{eq:boundx0}
  \end{eqnarray}

Multiplying Eq. \!(\ref{eq:ampl}) by $\frac{dA}{dx}$ and using Eqs. \!(\ref{eq:centerd}) and (\ref{eq:center}), Eq. \!(\ref{eq:ampl}) can be integrated as
  \begin{eqnarray}
  \left(A\frac{dA}{dx}\right)^2=\frac{1}{\xi_0^2}(1-A^2)
  \left(-(Q\xi_0)^2+\left(\frac{2\mu}{gn_0}-1\right)A^2-A^4\right).
  \label{eq:foo}  
  \end{eqnarray}
Here, we introduced the healing length $\xi_0\equiv(mgn_0)^{-\frac{1}{2}}$.
Integrating Eq. \!(\ref{eq:foo}) again~\cite{rf:snbook}, the solution of Eq. \!(\ref{eq:ampl}) in the region $\left(j-\frac{1}{2}\right)a<x<\left(j+\frac{1}{2}\right)a$ is given by 
 \begin{eqnarray}
   A(x)^2=(1-\beta_{-}){\rm sn}^2\left(\frac{\sqrt{\beta_{+}-\beta_{-}}
       (|x-ja|+x_0)}{\xi_0},
        \sqrt{\frac{1-\beta_{-}}{\beta_{+}-\beta_{-}}}\right)+\beta_{-},
        \label{eq:dens}
 \end{eqnarray}
%
where
  \begin{eqnarray}
  \beta_{\pm}\equiv 
  \frac{\mu}{gn_0}-\frac{1}{2}
  \pm\frac{1}{2}\sqrt{\left(\frac{2\mu}{gn_0}-1\right)^2-4(Q\xi_0)^2}.
  \end{eqnarray}
$\mu$ and $x_0$ are determined as functions of $V_0$, $a$ and $Q$ by solving Eq. \!(\ref{eq:center}) and Eq. \!(\ref{eq:boundx0}).
Once the amplitude $A(x)$ is obtained, the phase $S(x)$ can be calculated as
  \begin{eqnarray}
  S(x)-S(ja)=\int_{ja}^{x}dx\frac{Q}{A^2}.
  \label{eq:phasedef}
  \end{eqnarray}
The exact solution of Eq. \!(\ref{eq:dens}) and \!(\ref{eq:phasedef}) has been obtained in Refs.~\cite{rf:seaman1,rf:witthaut}.
The exact solution in a unit cell for $(a,V_0,Q)=(15 \,\xi_0,5 \,gn_0\xi_0,0.09 \,\xi_0^{-1})$ is shown in Figs. \ref{fig:ampl} and \ref{fig:phase} with dashed lines. From the exact solution, $\mu$ and $x_0$ are given by $(\mu,x_0)=(1.0040520 \,gn_0, 0.1364985 \,\xi_0)$.

For further analytic calculations, we assume that the lattice constant is much larger than the healing length. The condensate wave function far from each potential barrier asymptotically approaches the wave function of a uniform condensate.
Accordingly, the chemical potential is given by~\cite{rf:fetter}
  \begin{eqnarray}
  \mu=gn_0+\epsilon_Q,\label{eq:chem_homo}
  \end{eqnarray}
where $\epsilon_Q\equiv \frac{Q^2}{2m}$.
Substituting Eq. \!(\ref{eq:chem_homo}) into Eq. \!(\ref{eq:dens}), an approximate solution of Eq. \!(\ref{eq:ampl}) in the region $\left(j-\frac{1}{2}\right)a<x<\left(j+\frac{1}{2}\right)a$ can be obtained as
  \begin{eqnarray}
   A(x)^2=\gamma(x)^2+(Q\xi_0)^2,
   \label{eq:dens_th}
  \end{eqnarray}
where
  \begin{eqnarray}
  \gamma(x)\equiv\sqrt{1-(Q\xi_0)^2}
  {\rm tanh}\left(\frac{\sqrt{1-(Q\xi_0)^2}(|x-ja|+x_0)}{\xi_0}\right).\label{eq:gamma}
  \end{eqnarray}
Substituting Eq. \!(\ref{eq:dens_th}) into Eq. \!(\ref{eq:phasedef}), the phase $S(x)$ in the region $\left(j-\frac{1}{2}\right)a<x<\left(j+\frac{1}{2}\right)a$ is given by
  \begin{eqnarray}
   S(x)-S(ja)
   &=&\int_{ja}^{x}dx\, \frac{Q}{A^2}
   \nonumber\\
   &=&Q(x-ja)+{\rm sgn}(x) 
   \Biggl[{\rm tan}^{-1}\left(\frac{\gamma(x)}{Q\xi_0}\right)
   \nonumber\\
   & &-\mathrm{tan}^{-1}\left(\frac{\gamma(ja)}{Q\xi_0}\right)\Biggr].
   \label{eq:phase_th}
   \end{eqnarray}
From Eqs. \!(\ref{eq:dens_th}) and (\ref{eq:phase_th}), the condensate wave function is expressed as
  \begin{eqnarray}
  \Psi_0(x)=\sqrt{n_0}{\rm e}^{{\rm i}\left(Qx-{\rm sgn}(x)\,\theta_0\right)}
            \left(\gamma(x)-{\rm sgn}(x)\,{\rm i}Q\xi_0\right),
  \label{eq:con_cur}
  \end{eqnarray}
where
  \begin{eqnarray}
  {\rm e}^{{\rm i}\theta_0}\equiv
  \frac{\gamma(ja)-{\rm i}Q\xi_0}{\sqrt{\gamma(ja)^2+(Q\xi_0)^2}}.
  \end{eqnarray}
The wave function Eq. (\ref{eq:con_cur}) takes the similar form with that of a gray soliton~\cite{rf:soliton,rf:tsuzuki}.
The only difference between them is the constant $x_0$ in Eq. (\ref{eq:gamma}) which depends on the potential strength $V_0$.
Substituting Eq. \!(\ref{eq:con_cur}) into Eq. \!(\ref{eq:boundx0}), the condition for obtaining $x_0$ as a function of $Q$ and $V_0$ is 
  \begin{eqnarray} 
  \gamma(ja)^3+\frac{V_0}{gn_0\xi_0}\gamma(ja)^2
  -\left(1-(Q\xi_0)^2\right)\gamma(ja)
  +\frac{V_0}{gn_0\xi_0}(Q\xi_0)^2=0.\label{eq:gamma}
  \end{eqnarray}
The approximate solution of the amplitude Eq. \!(\ref{eq:dens_th}) and phase Eq. \!(\ref{eq:phase_th}) in a unit cell are shown in Figs. \ref{fig:ampl} and \ref{fig:phase} with solid lines. $V_0$ and $Q$ are chosen as $(V_0,Q)=(5\,gn_0\xi_0,0.09 \,\xi_0^{-1})$. One sees that the exact and approximate solutions show a good agreement when $a \gg \xi_0$.
From the approximate solution, $\mu$ and $x_0$ can be obtained as $(\mu,x_0)=(1.0040500 \,gn_0, 0.1364977 \,\xi_0)$ which are very close to the values calculated from the exact solultion.
We will also apply this approximation to the calculation of the excitation spectrum in Sec. \!IV.
Note that Eqs. \!(\ref{eq:dens}) and (\ref{eq:dens_th}) coincide with Eqs. \!(11) and (15) in Ref.~\cite{rf:wareware} when $Q=0$ and $x_0\neq 0$.

\subsection{Condensate band, group velocity and effective mass}

According to Refs.~\cite{rf:ware,rf:baratoff}, the phase difference $\varphi$ between the condensates in the neighboring sites is defined by
  \begin{eqnarray}
  \varphi\equiv Q\int_{(j-1/2)a}^{(j+1/2)a}dx
  \left(\frac{1}{A^2}-1\right).\label{eq:phd}
  \end{eqnarray}
Substituting Eq. \!(\ref{eq:phase_th}) into Eq. \!(\ref{eq:phd}), one obtains
  \begin{eqnarray}
  \varphi=2\left[{\rm tan}^{-1}\left(\frac{\sqrt{1-(Q\xi_0)^2}}{Q\xi_0}\right)
   -\mathrm{tan}^{-1}\left(\frac{\gamma(ja)}{Q\xi_0}\right)\right].
   \label{eq:fai}
  \end{eqnarray}
$\varphi$ can be interpreted as a phase jump across the potential barrier as shown in Fig. \ref{fig:phase}. $\varphi$ and $K$ are easily related by the Bloch theorem.

\begin{figure}[tb]
\includegraphics[width=7.0cm,height=4.5cm]{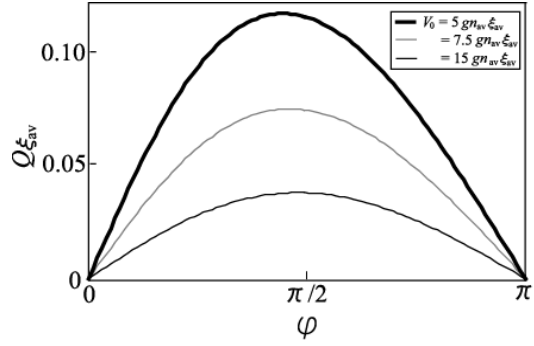}
\caption{\label{fig:curQ}
Superfluid momentum $Q$ as a function of the phase difference $\varphi$.
The thick black line, the gray line and the thin black line represent the superfluid momenta with $(a,V_0)=(15\,\xi_{\rm av},5\,gn_{\rm av}\xi_{\rm av})$, $(15\,\xi_{\rm av},7.5\,gn_{\rm av}\xi_{\rm av})$ and $(15\,\xi_{\rm av},15\,gn_{\rm av}\xi_{\rm av})$, respectively.
}
\end{figure}
Substituting Eq. \!(\ref{eq:con_cur}) into the normalization condition Eq. \!(\ref{eq:normal}), one obtains
  \begin{eqnarray}
  1-\frac{2\xi_0}{a}\sqrt{1-(Q\xi_0)^2}+\frac{2\xi_0}{a}\gamma(ja)
  = \frac{n_{\rm av}}{n_0},\label{eq:nave}
  \end{eqnarray}
where $n_{\rm av}\equiv\frac{N_{\rm C}}{a}$ is the average density. The healing length for the average density $n_{\rm av}$ is defined by $\xi_{\rm av}\equiv(mgn_{\rm av})^{-\frac{1}{2}}$.

Equations (\ref{eq:gamma}), (\ref{eq:fai}) and (\ref{eq:nave}) yield $Q$, $\gamma(ja)$ and $\mu$ as functions of $a$, $V_0$ and $\varphi$.
If Eqs. (\ref{eq:gamma}), (\ref{eq:fai}) and (\ref{eq:nave}) are expanded into the power series of $gn_{\rm av}\xi_{\rm av}/V_0$ and $\xi_{\rm av}/a$ assuming $V_0\gg gn_{\rm av}\xi_{\rm av}$ and $a\gg \xi_{\rm av}$, one obtains
  \begin{eqnarray}
  Q &\simeq& \frac{gn_{\rm av}}{2V_0}{\rm sin}\varphi,
  \label{eq:curph}
  \end{eqnarray}
  \begin{eqnarray}
  \gamma(ja) &\simeq& 
  \frac{gn_{\rm av}\xi_{\rm av}}{2V_0}(1+{\rm cos}\varphi),
  \label{eq:gamph}
  \end{eqnarray}
  \begin{eqnarray}
  \frac{\mu}{gn_{\rm av}}&\simeq& \frac{\mu|_{\varphi=0}}{gn_{\rm av}}
  +\frac{gn_{\rm av}\xi_{\rm av}^2}{a V_0}(1-{\rm cos}\varphi)
  +\frac{(gn_{\rm av}\xi_{\rm av})^2}{8V_0^2}{\rm sin}^2\varphi,
  \label{eq:chemph}\\
  \frac{\mu|_{\varphi=0}}{gn_{\rm av}}&\simeq&1+\frac{2\xi_{\rm av}}{a}
  +\frac{2\xi_{\rm av}^2}{a^2}-\frac{2gn_{\rm av}\xi_{\rm av}^2}{a V_0}.
  \end{eqnarray}
Substituting Eqs. \!(\ref{eq:con_cur}), (\ref{eq:curph}), (\ref{eq:gamph}) and (\ref{eq:chemph}) into Eq. \!(\ref{eq:mean_ener}), the energy of the condensate is
  \begin{eqnarray}
  \frac{E}{N_{\rm C}gn_{\rm av}}&\simeq&
  \frac{E|_{\varphi=0}}{N_{\rm C}gn_{\rm av}}
  +\frac{gn_{\rm av}\xi_{\rm av}^2}{2a V_0}(1-{\rm cos}\varphi)
  +\frac{(gn_{\rm av}\xi_{\rm av})^2}{8V_0^2}{\rm sin}^2\varphi,
  \label{eq:enerph}\\
  \frac{E|_{\varphi=0}}{N_{\rm C}gn_{\rm av}}&\simeq&
  \frac{1}{2}+\frac{4\xi_{\rm av}}{3a}
  +\frac{2\xi_{\rm av}^2}{a^2}-\frac{gn_{\rm av}\xi_{\rm av}^2}{a V_0}.
  \end{eqnarray}
We have rescaled all the variables by the averaged density $n_{\rm av}$ instead of the center density $n_0$, because $n_{\rm av}$ ($n_0$) is independent (dependent) of the condensate quasimomentum $K$. The higher order terms of $gn_{\rm av}\xi_{\rm av}/V_0$ and $\xi_{\rm av}/a$ are not shown just for avoiding the complication of the equations.
The higher order terms of the expansion are shown in Appendix, because they are necessary for the calculations in the following sections.

In Fig. \ref{fig:curQ}, $Q$ is shown as a function of $\varphi$.
In Eq. \!(\ref{eq:curph}) and Fig. \ref{fig:curQ}, we can clearly see the well-known Josephson relation between the supercurrent and phase difference  in the limit of $V_0\gg gn_{\rm av}\xi_{\rm av}$~\cite{rf:ware}. 

%
\begin{figure}[tb]
\includegraphics[width=7.5cm,height=5.0cm]{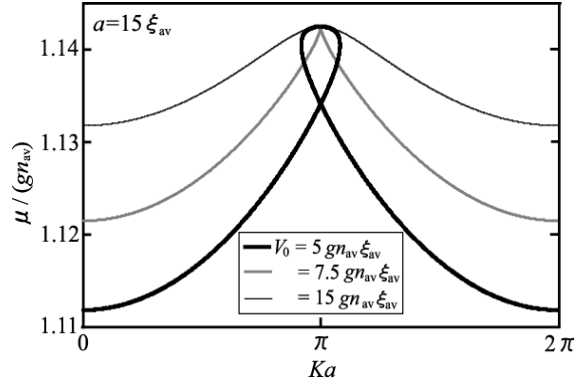}
\caption{\label{fig:chem}
Chemical potential $\mu$ as a function of the quasimomentum $K$.
The thick black line, the gray line and the thin black line represent the chemical potentials with $(a,V_0)=(15\,\xi_{\rm av},5\,gn_{\rm av}\xi_{\rm av})$, $(15\,\xi_{\rm av},7.5\,gn_{\rm av}\xi_{\rm av})$ and $(15\,\xi_{\rm av},15\,gn_{\rm av}\xi_{\rm av})$, respectively.
}
\end{figure}
\begin{figure}[tb]
\includegraphics[width=8cm, height=5cm]{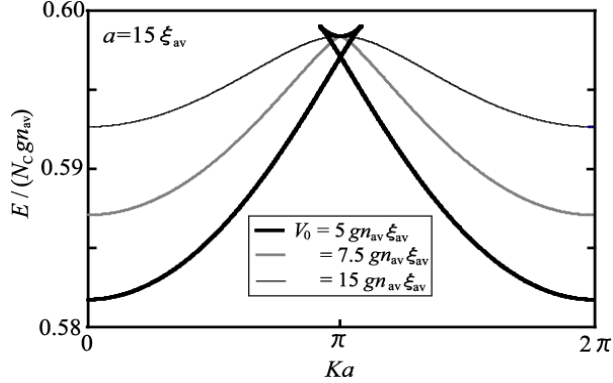}
\caption{\label{fig:meanener}
Energy $E$ of the condensate per site as a function of the quasimomentum $K$, namely the first condensate band.
The thick black line, the gray line and the thin black line represent the first condensate bands with $(a,V_0)=(15\,\xi_{\rm av},5\,gn_{\rm av}\xi_{\rm av})$, $(15\,\xi_{\rm av},7.5\,gn_{\rm av}\xi_{\rm av})$ and $(15\,\xi_{\rm av},15\,gn_{\rm av}\xi_{\rm av})$, respectively.
}
\end{figure}

Imposing the Bloch theorem
  \begin{eqnarray}
  \Psi_0(x+a)=\Psi_0(x){\rm e}^{{\rm i}Ka}\label{eq:bloch_con}
  \end{eqnarray}
on the condensate wave function Eq. \!(\ref{eq:con_cur}), the relation between $K$ and $\varphi$ is derived as
  \begin{eqnarray}
  Ka=Qa+\varphi.\label{eq:blochcon}
  \end{eqnarray}
Substituting Eq. \!(\ref{eq:blochcon}) into Eqs. \!(\ref{eq:chemph}) and (\ref{eq:enerph}), $\mu(K)$ and $E(K)$ can be calculated as shown in Figs. \ref{fig:chem} and \ref{fig:meanener}. A loop structure of $\mu(K)$ and a swallow-tail of $E(K)$ are present at the edge of the first Brillouin zone for relatively small $V_0$ (thick black lines).  The loop structure and swallow-tail become smaller as $V_0$ increases and they disappear for a certain critical value of $V_0$ (gray lines).
The condition for the presence of swallow-tail (loop structure) is obtained from Eq. \!(\ref{eq:blochcon}) as follows.
A swallow-tail is present if there are two values of $\varphi$ which satisfy Eq. \!(\ref{eq:blochcon}) with $Ka=\pi$, while a swallow-tail is not present if there is only one value of $\varphi$ which satisfies Eq. \!(\ref{eq:blochcon}) with $Ka=\pi$.
This condition yields the threshold
  \begin{eqnarray}
  V_0=\frac{gn_{\rm av} a}{2},\label{eq:swallowcond}
  \end{eqnarray}
which agrees with the result obtained by the numerical calculations in Ref.~\cite{rf:seaman1}.

\begin{figure}[tb]
\includegraphics[width=7.5cm,height=5.5cm]{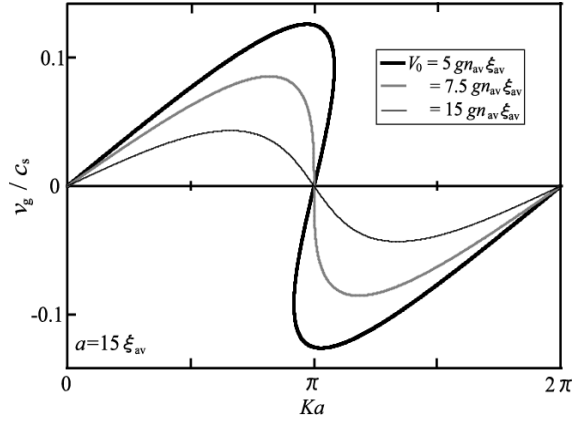}
\caption{\label{fig:gvel}
Group velocity $v_{\rm g}$ as a function of the quasimomentum $K$.
The thick black line, the gray line and the thin black line represent the group velocities with $(a,V_0)=(15\,\xi_{\rm av},5\,gn_{\rm av}\xi_{\rm av})$, $(15\,\xi_{\rm av},7.5\,gn_{\rm av}\xi_{\rm av})$ and $(15\,\xi_{\rm av},15\,gn_{\rm av}\xi_{\rm av})$, respectively.
}
\end{figure}

The first derivative of the condensate band gives the group velocity,
  \begin{eqnarray}
  v_{\rm g}=\frac{\partial E}{\partial K}.
  \label{eq:gvelph}
  \end{eqnarray}
Assuming $V_0\gg gn_{\rm av}\xi_{\rm av}$ and substituting Eq. \!(\ref{eq:enerph}) into Eq. \!(\ref{eq:gvelph}), $v_{\rm g}$ is given by
  \begin{eqnarray}
  \frac{v_{\rm g}}{c_{\rm s}}\simeq 
  \frac{gn_{\rm av}\xi_{\rm av}}{2V_0}{\rm sin}\varphi,
  \label{eq:gvelphap}
  \end{eqnarray}
where $c_{\rm s}\equiv\sqrt{\frac{gn_{\rm av}}{m}}$ is the sound velocity for a uniform condensate.
Substituting Eq. \!(\ref{eq:blochcon}) into Eq. \!(\ref{eq:gvelphap}), $v_{\rm g}(K)$ is obtained as shown in Fig. \ref{fig:gvel}.
When $V_0$ is small, $v_{\rm g}(K)$ has a reentrant structure reflecting the presence of the swallow-tail (thick black line).
When the potential strength is larger than the critical value of Eq. \!(\ref{eq:swallowcond}), the reentrant structure is absent.

In the limit of $V_0\gg gn_{\rm av} a$, the first condensate band and group velocity take the form of the tight-binding approximation~\cite{rf:wuniu2,rf:smerzi1,rf:smerzi2},
  \begin{eqnarray}
  \frac{E}{N_{\rm C}gn_{\rm av}} &\simeq& 
  \frac{E|_{\varphi=0}}{N_{\rm C}gn_{\rm av}}
                   +\frac{gn_{\rm av}\xi_{\rm av}^2}{2a V_0}(1-{\rm cos}Ka),\\
  \frac{v_{\rm g}}{c_{\rm s}}
  &\simeq& \frac{gn_{\rm av}\xi_{\rm av}}{2V_0} \sin Ka.
  \label{eq:tightv}
  \end{eqnarray}

\begin{figure}[tb]
\includegraphics[width=7.5cm,height=5.0cm]{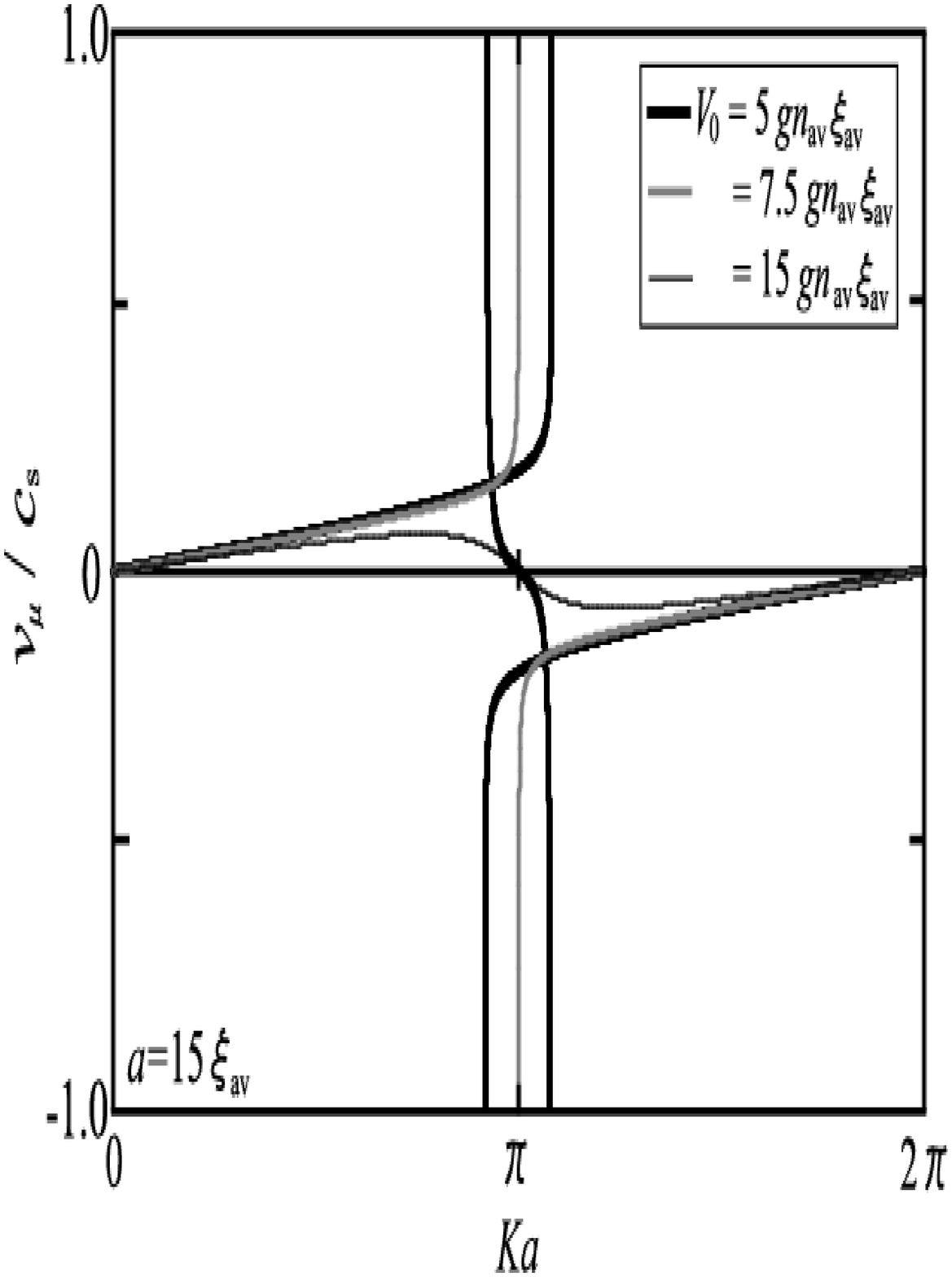}
\caption{\label{fig:chvel}
Group velocity $v_{\mu}$ as a function of the quasimomentum $K$.
The thick black line, the gray line and the thin black line represent the group velocities with $(a,V_0)=(15\,\xi_{\rm av},5\,gn_{\rm av}\xi_{\rm av})$, $(15\,\xi_{\rm av},7.5\,gn_{\rm av}\xi_{\rm av})$ and $(15\,\xi_{\rm av},15\,gn_{\rm av}\xi_{\rm av})$, respectively.
}
\end{figure}

One can define another type of group velocity by
  \begin{eqnarray}
  v_{\mu}=
  \frac{\partial\mu}{\partial K}.
  \label{eq:gchph}
  \end{eqnarray}
Assuming $V_0\gg gn_{\rm av}\xi_{\rm av}$ and substituting Eq. \!(\ref{eq:chemph}) into Eq. \!(\ref{eq:gchph}), $v_{\mu}$ is given by
  \begin{eqnarray}
  \frac{v_{\mu}}{c_{\rm s}}
  &\simeq&
  \frac{
  \frac{gn_{\rm av}}{V_0}{\rm sin}\varphi
  \left(\frac{\xi_{\rm av}}{a}
  +\frac{gn_{\rm av}\xi_{\rm av}}{4V_0}{\rm cos}\varphi\right)
  }
  {
  \frac{\partial K}{\partial\varphi}
  },
  \label{eq:gchphap}
  \end{eqnarray}
  \begin{eqnarray}
  \frac{\partial K}{\partial\varphi}\simeq
  \left(\frac{1}{a}
  +\frac{gn_{\rm av}}{2V_0}{\rm cos}\varphi\right).
  \end{eqnarray}
Substituting Eq. \!(\ref{eq:blochcon}) into Eq. \!(\ref{eq:gchphap}), $v_{\mu}(K)$ is obtained as shown in Fig. \ref{fig:chvel}. $v_{\mu}$ diverges at the edge of the swallow-tail (thick black line).

It is important to remark about the physical meaning of $v_{\rm g}$ and $v_{\mu}$. $v_{\rm g}$ can be regarded as a superfluid velocity of a condensate flowing through a periodic potential, because $n_{\rm av}v_{\rm g}$ is equal to the supercurrent $\frac{n_0 Q}{m}$. 
On the other hand, the velocity of Bogoliubov phonons propagating in the opposite (same) direction to the supercurrent reduces (increases) by $v_{\mu}$ and that the Landau instability occurs when $v_{\mu}$ exceeds the Bogoliubov sound velocity~\cite{rf:pesmith,rf:edo,rf:smerzi2,rf:konabe2}.
We will discuss the detail of this issue in Sec. V.

\begin{figure}[tb]
\includegraphics[width=7.5cm,height=5.0cm]{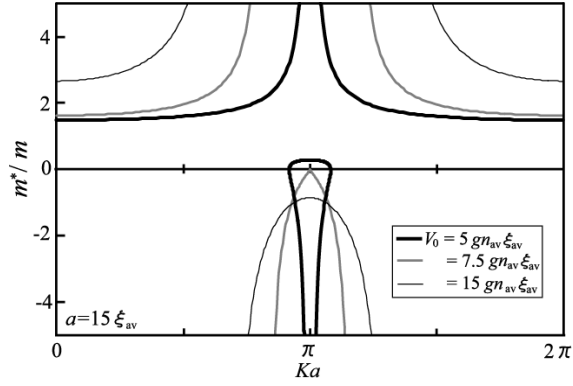}
\caption{\label{fig:effmas}
Effective mass $m^{\ast}$.
The thick black line, the gray line and the thin black line represent the effective masses with $(a,V_0)=(15\,\xi_{\rm av},5\,gn_{\rm av}\xi_{\rm av})$, $(15\,\xi_{\rm av},7.5\,gn_{\rm av}\xi_{\rm av})$ and $(15\,\xi_{\rm av},15\,gn_{\rm av}\xi_{\rm av})$, respectively.
}
\end{figure}
The effective mass is defined by the inverse of the second derivative of the condensate band as
  \begin{eqnarray}
  m^{\ast}=
  \left(\frac{\partial^2 E}{\partial K^2}\right)^{-1}=\left(\frac{\partial v_{\rm g}}{\partial K}\right)^{-1}.
  \label{eq:effmph}
  \end{eqnarray}
Assuming $V_0\gg gn_{\rm av}\xi_{\rm av}$ and substituting Eq. \!(\ref{eq:enerph}) into Eq. \!(\ref{eq:effmph}), $m^{\ast}$ is given by
  \begin{eqnarray}
  \frac{m^{\ast}}{m}\simeq
  \frac{2V_0}{gn_{\rm av}{\rm cos}\varphi}\frac{\partial K}{\partial \varphi}.
  \label{eq:effmphap}
  \end{eqnarray}
Substituting Eq. \!(\ref{eq:blochcon}) into Eq. \!(\ref{eq:effmphap}), $m^\ast(K)$ is obtained as shown in Fig. \ref{fig:effmas}.
The sign of the effective mass changes when the group velocity takes its extremum.
$m^\ast(K)$ also changes its sign at the edge of the swallow-tail, because $\frac{\partial v_{\rm g}}{\partial K}$ diverges there. $m^\ast(K)$ is always positive in the upper portion of the swallow-tail and negative near the edge of the swallow-tail in the lower portion.
This is consistent with the result of the exact solution Eqs. \!(\ref{eq:dens}) and (\ref{eq:phasedef})~\cite{rf:prep}.

Since the sound velocity in a periodic potential is given by~\cite{rf:menotti}
  \begin{eqnarray}
  c_{\rm b}=\sqrt{\frac{n_{\rm av}}{m^{\ast}}
                  \frac{\partial\mu}{\partial n_{\rm av}}},
  \label{eq:bogophonon}
  \end{eqnarray}
the negative effective mass leads to the dynamical instability due to long-wavelength phonons~\cite{rf:pesmith,rf:edo,rf:smerzi2,rf:seaman1}, 
and the formation of bright gap solitons follows after the dynamical instability~\cite{rf:gapsol,rf:konotop}.
Condensates are expected to be stable for the long-wavelength perturbation in the upper portion of a swallow-tail, 
because of the positive effective mass.
However, the upper portion of a swallow-tail corresponds to energy saddle points in the two-state approximation and is predicted to be unstable \cite{rf:mueller}. The dynamical instability of the upper portion of a swallow-tail has been shown by numerically solving the time-dependent GP equation~\cite{rf:seaman1}.
We will calculate the excitation spectrum in the following sections and show that the spectrum has a phonon branch but also has a different branch causing the dynamical instability.


\begin{figure}[tb]
\includegraphics[width=7.0cm,height=4.5cm]{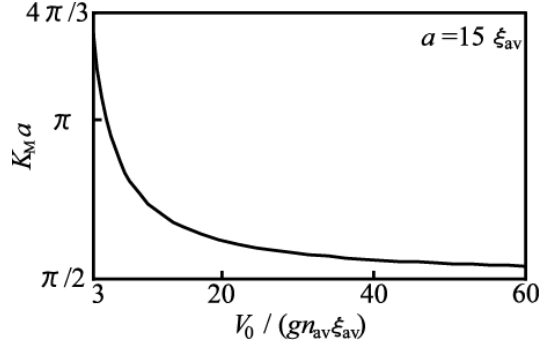}
\caption{\label{fig:kmaxv}
Condensate quasimomentum $K_{\rm M}$ which gives the maximum group velocity
}
\end{figure}
\begin{figure}[tb]
\includegraphics[width=7.0cm,height=4.5cm]{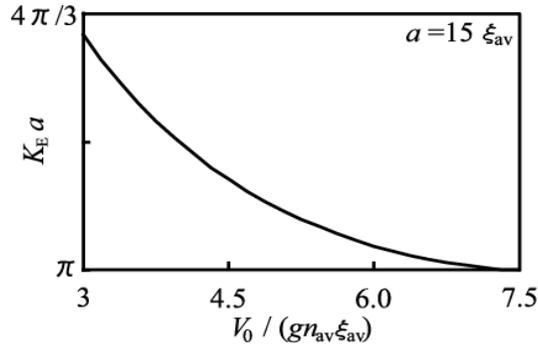}
\caption{\label{fig:kedge}
Condensate quasimomentum $K_{\rm E}$ which gives the edge of the swallow-tail.
}
\end{figure}

From Fig. 9, it can be seen that the effective mass changes its sign twice for positive $K$.
Since the sign of the effective mass is closely related to the stability of condensates, it is important to calculate the quasimomentum at which the effective mass changes its sign.
$K_{\rm M}$ is determined by the condition of maximum group velocity as
  \begin{eqnarray}
  \frac{\partial v_{\rm g}}{\partial K}=0.
  \label{eq:condi_mv}
  \end{eqnarray}
Assuming $V_0\gg gn_{\rm av}\xi_{\rm av}$, the phase difference $\varphi_{\rm M}$ when $K=K_{\rm M}$ can be calculated from Eqs. \!(\ref{eq:condi_mv}) and (\ref{eq:vgho}) in Appendix as
  \begin{eqnarray}
  {\rm cos}\varphi_{\rm M}= \frac{gn_{\rm av}\xi_{\rm av}}{V_0}
  -\frac{gn_{\rm av}\xi_{\rm av}^2}{a V_0}
  -\frac{(gn_{\rm av}\xi_{\rm av})^2}{2V_0^2}.
  \label{eq:phimax}
  \end{eqnarray}
Substituting Eq. \!(\ref{eq:phimax}) into Eq. \!(\ref{eq:blochcon}), $K_{\rm M}$ is obtained as shown in Fig. \ref{fig:kmaxv}.
$K_{\rm M}a$ asymptotically approaches $\frac{\pi}{2}$ in the limit of $V_0 \gg gn_{\rm av}a$, because the group velocity takes the form of Eq. \!(\ref{eq:tightv}).

The effective mass changes its sign at the edge of the swallow-tail $K=K_{\rm E}$. $K_{\rm E}$ is determined by the condition,
  \begin{eqnarray}
   m^{\ast}=0.
   \label{eq:condi_edge}
  \end{eqnarray}
Assuming $V_0\gg gn_{\rm av}\xi_{\rm av}$, the phase difference $\varphi_{\rm E}$ when $K=K_{\rm E}$ is obtained by Eqs. \!(\ref{eq:condi_edge}) and (\ref{eq:effmho}) in Appendix,
  \begin{eqnarray}
  {\rm cos}\varphi_{\rm E}=-\frac{2V_0}{gn_{\rm av}a}
  +\frac{gn_{\rm av}\xi_{\rm av}}{V_0}-\frac{4V_0\xi_{\rm av}}{gn_{\rm av}a^2}
  -\frac{(gn_{\rm av}\xi_{\rm av})^2}{2V_0^2}
  +\frac{2\xi_{\rm av}^2}{a^2}.
  \label{eq:phiE}
  \end{eqnarray}
Substituting Eq. \!(\ref{eq:phiE}) into Eq. \!(\ref{eq:blochcon}), $K_{\rm E}$ is obtained as shown in Fig. \ref{fig:kedge}.
The size of the swallow-tail monotonically decreases until the potential strength reaches the critical value $V_0=\frac{gn_{\rm av}a}{2}$.

\section{Bogoliubov excitation spectrum}
In this section, we shall solve the Bogoliubov equations and calculate the excitation spectrum of a current-carrying condensate in a KP potential.
We hereafter call the band structure of excitation spectrum as the {\it Bogoliubov band}.
As is well known, the band structure of a single particle in a periodic potential can be calculated from a single barrier problem~\cite{rf:ashcroft}.
In our previous work, the Bogoliubov band of a current-free condensate in a KP potential has been calculated from a single-barrier problem for Bogoliubov excitations~\cite{rf:wareware}.
We shall extend this work to the case of current-carrying condensates.
\subsection{Single-barrier problem}
In a recent work, the scattering problem of Bogoliubov excitations of a current-carrying condensate has been studied, and it has been found that the tunneling properties, such as transmission coefficient and phase shift, strongly depend on the presence of a supercurrent~\cite{rf:ware}.
Here, we briefly review the calculation in Ref. \cite{rf:ware}, since it is necessary for the calculation of the Bogoliubov band.

We assume a single potential barrier $V(x)=V_0\delta(x)$. The solution of the time-independent GP equation is given by Eq. \!(\ref{eq:con_cur}). 
The solution of the Bogoliubov equations with Eq. \!(\ref{eq:con_cur}) can be calculated from the solution for a gray soliton obtained in Ref.~\cite{rf:soliton}.
Substituting Eq. \!(\ref{eq:con_cur}) into Eq. (\ref{eq:BdGE}), one can obtain the four particular solutions
 \begin{eqnarray}
  u_n(x) &=&
  \Lambda_n {\rm e}^{{\rm i}[(k_n+Q) x
                     +{\rm sgn}(x)\frac{\varphi}{2}]}
     \Biggl\{\Biggr.
     \left(1+\frac{(k_n\xi_0)^2 gn_0}{2\varepsilon}\right)\gamma(x)
     -{\rm i}\,{\rm sgn}(x)
       \nonumber \\
  && \times\left[Q\xi_0
     +\frac{k_n \xi_0 gn_0}{2\varepsilon}\left(1-(Q\xi_0)^2-\gamma(x)^2
     +\frac{\varepsilon}{gn_0}\right)
     +\frac{(k_n\xi_0)^3 gn_0}{4\varepsilon}\right]
     \Biggl.\Biggr\},
     \label{eq:anslt}\\
  v_n(x) &=& 
  \Lambda_n {\rm e}^{{\rm i}[(k_n-Q) x
                     -{\rm sgn}(x)\frac{\varphi}{2}]}
     \Biggl\{\Biggr.
     \left(1-\frac{(k_n\xi_0)^2 gn_0}{2\varepsilon}\right)\gamma(x)
     +{\rm i}\,{\rm sgn}(x)
       \nonumber \\
  && \times\left[Q\xi_0+\frac{k_n\xi_0 gn_0}{2\varepsilon}
     \left(1-(Q\xi_0)^2-\gamma(x)^2-\frac{\varepsilon}{gn_0}\right)
     +\frac{(k_n\xi_0)^3 gn_0}{4\varepsilon}\right]
     \Biggl.\Biggr\},
 \end{eqnarray}
 for $1\le n \le 4$.
$k_n$ is a solution of
  \begin{eqnarray}
  \varepsilon=\frac{Q k_n}{m}
              +\sqrt{\epsilon_{k_n}(\epsilon_{k_n}+2gn_0)},
                    \label{eq:bg_sp}
  \end{eqnarray}
for a given $\varepsilon$, where $\epsilon_{k_n}\equiv \frac{k_n^2}{2m}$.
Equation (\ref{eq:bg_sp}) is the Bogoliubov spectrum in a uniform system when the condensate flows with a velocity $\frac{Q}{m}$~\cite{rf:fetter}.
$\left(u_1(x), v_1(x)\right)$ and $\left(u_2(x), v_2(x)\right)$ describe the scattering components. $\left(u_3(x<0), v_3(x<0)\right)$ and $\left(u_4(x>0), v_4(x>0)\right)$ describe the localized components around the potential barrier. $\left(u_4(x<0), v_4(x<0)\right)$ and $\left(u_3(x>0), v_3(x>0)\right)$ describe the components which diverge for $|x|\to \infty$~\cite{rf:ware}.
The normalization constant $\Lambda_n$ is given by
  \begin{eqnarray}
  \Lambda_n &=& \left\{\!\begin{array}{cc}
  \frac{{\rm e}^{{\rm i}\alpha_1}}
  {\sqrt{2\left(\frac{\varepsilon}{gn_0}-Q k_1\xi_0^2\right)}}\,\,\,n=1, 3, 4\\
  \frac{{\rm e}^{{\rm i}\alpha_2}}
  {\sqrt{2\left(\frac{\varepsilon}{gn_0}-Q k_2\xi_0^2\right)}}, \,\,\, n=2
                  \end{array}\!\right.,
  \end{eqnarray}
where
  \begin{eqnarray}
  {\rm e}^{{\rm i}\alpha_n}\equiv&&
  \frac{2gn_0\varepsilon+\varepsilon \frac{Q k_n}{m}
  +2(gn_0-2\epsilon_Q)\epsilon_{k_n}+\epsilon_{k_n}\frac{Qk_n}{m}}
  {2\varepsilon\sqrt{gn_0\left(gn_0+\epsilon_{k_n}
                           +\varepsilon-\frac{Q k_n}{m}\right)}}
  \nonumber\\
  &&+{\rm i}\frac{gn_0\sqrt{1-(Q\xi_0)^2}k_n\xi_0
  (\varepsilon-\frac{Qk_n}{m}+\epsilon_{k_n})}
  {2\varepsilon\sqrt{gn_0\left(gn_0+\epsilon_{k_n}
                           +\varepsilon-\frac{Q k_n}{m}\right)}}.
  \end{eqnarray}
The normalization constants of the scattering components are determined to satisfy $\left(u_1(x), v_1(x)\right)\rightarrow\left(u_{k_1}, v_{k_1}\right)$ and $\left(u_2(x), v_2(x)\right)\rightarrow\left(u_{k_2}, v_{k_2}\right)$ for $x\to \infty$, where the amplitudes $u_{k}$ and $v_{k}$ are the well-known Bogoliubov transformation coefficients given by
  \begin{eqnarray}
  u_k=
  \sqrt{\frac{gn_0+\epsilon_k+\varepsilon-\frac{Q k}{m}}
  {2(\varepsilon-\frac{Q k}{m})}}
      {\rm e}^{{\rm i}(Q x+{\rm sgn}(x)\frac{\varphi}{2})},
  \\
  v_k=
  \sqrt{\frac{gn_0+\epsilon_k-\varepsilon+\frac{Q k}{m}}
  {2(\varepsilon-\frac{Qk}{m})}}
      {\rm e}^{-{\rm i}(Qx+{\rm sgn}(x)\frac{\varphi}{2})}.
  \end{eqnarray}
  
\begin{figure}[tb]
\includegraphics[width=9.5cm,height=4.5cm]{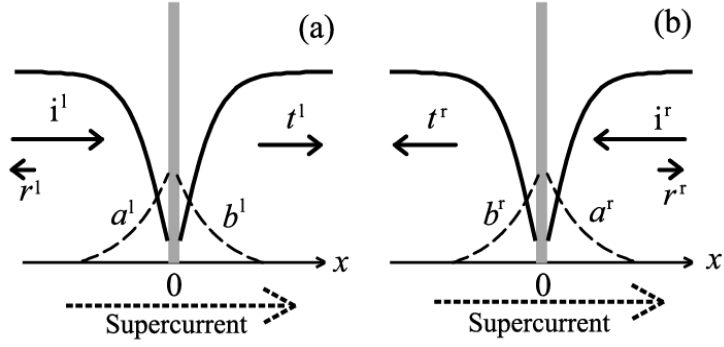}
\caption{\label{fig:tunnel_pic}
Schematic picture of the solutions of the single-barrier problem.
Scattering processes in which (a) the excitation comes from the left-hand side, and (b) the excitation comes from the right-hand side. The solid curves, solid arrows, dashed curves, and dashed arrows describe the condensate wave functions, scattering components, localized components and condensate supercurrent, respectively.
}
\end{figure}
Two independent eigenfunctions of Eq. \!(\ref{eq:BdGE}) corresponding to two types of scattering process are obtained by omitting divergent components. One is the process in which an excitation comes from the left-hand side (${\psi}^{\rm l}(x)$), and the other from the right-hand side (${\psi}^{\rm r}(x)$).
Schematic pictures of the solutions of the single-barrier problem are shown in Fig. \ref{fig:tunnel_pic}.
Here we consider the former written as
  \begin{eqnarray}
   \psi^{\rm l}(x)
   =
     \left(\!\begin{array}{cc}
                      u^{\rm l} \\ v^{\rm l}
                  \end{array}\!\right)
   \!\!\!&=&\!\!\!\left\{\begin{array}{ll}
          \left(\!\begin{array}{cc}
                    u_1 \\ v_1
                         \end{array}\!\right)
          +r^{\rm l}\left(\!\begin{array}{cc}
                    u_2 \\ v_2
                         \end{array}\!\right)
          +a^{\rm l}\left(\!\begin{array}{cc}
                    u_3 \\ v_3
                         \end{array}\!\right),
                         & x<0,  \\
          t^{\rm l}\left(\!\begin{array}{cc}
                    u_1 \\ v_1
                         \end{array}\!\right)
          +b^{\rm l}\left(\!\begin{array}{cc}
                    u_4 \\ v_4
                         \end{array}\!\right),
                         & x>0.
          \end{array}\right. \label{eq:lcs}
\end{eqnarray}
The coefficients $r^{\rm l}$, $a^{\rm l}$, $t^{\rm l}$, and $b^{\rm l}$ are the amplitudes of the reflected, left localized, transmitted, and right localized components, respectively. They are functions of the excitation energy $\varepsilon$, the potential strength $V_0$ and the phase difference of the condensate in the background $\varphi$.

The boundary conditions at $x=0$ are given by
  \begin{eqnarray}
     \psi^{\rm l}(+0)&=&\psi^{\rm l}(-0),\label{eq:bc2}\\
     \left. \frac{d{\psi}^{\rm l}}{dx} \right|_{+0}
     &=&\left. \frac{d\psi^{\rm l}}{dx} \right|_{-0}
     +2mV_0{\psi}^{\rm l}(0).
     \label{eq:bc}
  \end{eqnarray}
Equation (\ref{eq:bc}) can be obtained by integrating Eq. \!(\ref{eq:BdGE}) from $-0$ to $+0$.
The coefficients $r^{\rm l}$, $a^{\rm l}$, $t^{\rm l}$ and $b^{\rm l}$ are determined from Eqs. (\ref{eq:bc2}) and (\ref{eq:bc}).
The analytical expressions of the coefficients were obtained in Ref.~\cite{rf:ware} in the case of $\varepsilon \ll gn_0$ and $V_0 \gg gn_0\xi_0$.
Using the relations between the coefficients in ${\psi}^{\rm l}$ and ${\psi}^{\rm r}$
  \begin{eqnarray}
     \begin{array}{cc}
        r^{\rm l}(\varepsilon,-\varphi)=r^{\rm r}(\varepsilon,\varphi),\;
        a^{\rm l}(\varepsilon,-\varphi)=a^{\rm r}(\varepsilon,\varphi),
         \\
        t^{\rm l}(\varepsilon,-\varphi)=t^{\rm r}(\varepsilon,\varphi),\;
        b^{\rm l}(\varepsilon,-\varphi)=b^{\rm r}(\varepsilon,\varphi),
     \end{array}
     \label{eq:lr}
  \end{eqnarray}
one can also calculate ${\psi}^{\rm r}$.

\begin{figure}[tb]
      \includegraphics[width=7.0cm,height=4.5cm]{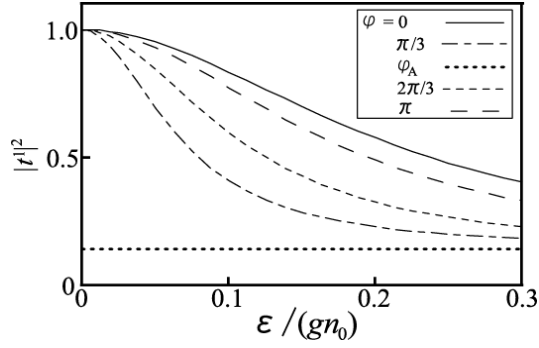}
\caption{\label{fig:antun_abs}
Transmission coefficient $T^{\rm l}$ as a function of $\varepsilon$ and $\varphi$ for $V_0=5 gn_0 \xi_0$.
$T^{\rm l}$s with $\varphi=0$, $\varphi=\pi/3$, $\varphi=\varphi_{\rm A}\simeq\pi/2$, $\varphi=2\pi/3$ and $\varphi=\pi$ are shown.
This figure is almost the same as Fig. 2(a) of Ref.~\cite{rf:ware} where $V_0=10 gn_0 \xi_0$.
}
\end{figure}
\begin{figure}[tb]
      \includegraphics[width=7.0cm,height=4.5cm]{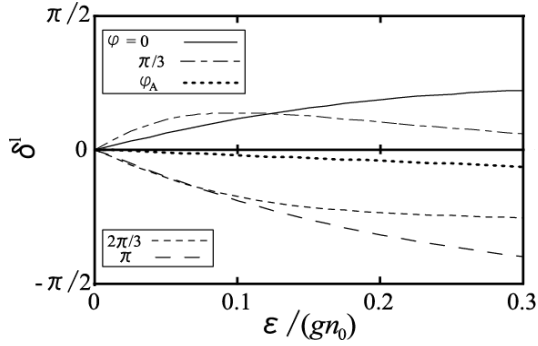}
\caption{\label{fig:antun_phs}
Phase shift $\delta^{\rm l}$ of $t^{\rm l}$ as a function of $\varepsilon$ and $\varphi$ for $V_0=5 gn_0\xi_0$.
$T^{\rm l}$s with $\varphi=0$, $\varphi=\pi/3$, $\varphi=\varphi_{\rm A}\simeq\pi/2$, $\varphi=2\pi/3$ and $\varphi=\pi$ are shown.
}
\end{figure}
In Figs. \ref{fig:antun_abs} and \ref{fig:antun_phs}, the transmission coefficient $T^{\rm l}\equiv|t^{\rm l}|^2$ and the phase shift $\delta^{\rm l}\equiv{\rm arg}(t^{\rm l})$ are shown as functions of the excitation energy $\varepsilon$ and the phase difference $\varphi$.
As shown in Ref.~\cite{rf:ware}, the transmission coefficient strongly depends on the phase difference as follows.
When $\varphi=0$, $T^{\rm l}$ approaches unity and $\delta^{\rm l}$ approaches zero as $\varepsilon$ approaches zero.
This means that the potential barrier is transparent for the low-energy excitations.
This behavior is called as {\it anomalous tunneling}~\cite{rf:kov,rf:antun}.
As $\varphi$ increases, the peak width of $T^{\rm l}$ decreases.
The peak width becomes zero and the sign of $\delta^{\rm l}$ changes when $\varphi=\varphi_{\rm A}\simeq\frac{\pi}{2}$. The anomalous tunneling is absent when $\varphi=\varphi_{\rm A}$.
As $\varphi$ increases further from $\varphi_{\rm A}$, the peak width increases and the anomalous tunneling appears again in the region $\varphi_{\rm A}<\varphi\leq \pi$.

The superfluid momentum $Q$ takes its maximum value when $\varphi=\varphi_{\rm A}$~\cite{rf:ware}.
Accordingly, $\varphi_{\rm A}$ is determined by the condition
  \begin{eqnarray}
  \frac{\partial Q}{\partial \varphi}=0.\label{eq:maxcur}
  \end{eqnarray}
When $V_0\gg gn_0\xi_0$, $\varphi_{\rm A}$ can be calculated from Eqs. \!(\ref{eq:maxcur}) and (\ref{eq:Qho}) (see appendix) as
  \begin{eqnarray}
  {\rm cos}\varphi_{\rm A} \simeq \frac{gn_{\rm av}\xi_{\rm av}}{V_0}
        +\frac{gn_{\rm av} \xi_{\rm av}^2}{aV_0}
        -\frac{(gn_{\rm av}\xi_{\rm av})^2}{2V_0^2}.
        \label{eq:phiano}
  \end{eqnarray}
Substituting Eq. \!(\ref{eq:phiano}) into Eq. \!(\ref{eq:blochcon}), $K_{\rm A}$ corresponding $\varphi_{\rm A}$ can be calculated.
Comparing Eqs. \!(\ref{eq:phiano}) and \!(\ref{eq:phimax}), one can see that $|K_{\rm A}|$, at which the anomalous tunneling is absent, is always smaller than $|K_{\rm M}|$, which gives the maximum supercurrent~\cite{rf:lastnote}.
As we will discuss in Sec. \!V, $K=K_{\rm A}$ is the onset of the Landau instability in a KP potential.

\subsection{Bogoliubov band}

Since the Bogoliubov equations are linear differential equations, a general solution of the equations can be described as a linear combination of independent solutions with the same energy. A general solution of the Bogoliubov equations with a single potential barrier can be written as a linear combination of $\psi^{\rm l}(x)$ and $\psi^{\rm r}(x)$
  \begin{eqnarray}
  \psi(x)=\alpha\psi^{\rm l}(x)+\beta\psi^{\rm r}(x),
  \label{eq:lcombi}
  \end{eqnarray}
where $\alpha$ and $\beta$ are arbitrary constants.

The solution of the Bogoliubov equations with the periodic potential Eq. (\ref{KPpotential}) can be calculated by imposing the Bloch theorem to Eq. \!(\ref{eq:lcombi}) as the current-free case in Ref. \cite{rf:wareware}. However, the non-diagonal element of Eq. (\ref{eq:BdGE}) does not have the periodicity of $a$ when the condensate has a supercurrent, because $\Psi_0(x)$ satisfies the Bloch theorem Eq. \!(\ref{eq:bloch_con}). Introducing $\Phi_0(x)$ and $\left(u_{\rm B}(x),v_{\rm B}(x)\right)$ by
  \begin{eqnarray}
  \Psi_0(x)&=&\Phi_0(x){\rm e}^{{\rm i}Kx},\\
  \left(u(x),v(x)\right)&=&\left(u_B(x){\rm e}^{{\rm i}Kx}
                            ,v_B(x){\rm e}^{-{\rm i}Kx}\right),
  \end{eqnarray}
  the Bogoliubov equations reduce to
       \begin{eqnarray}
               \left(
                 \begin{array}{cc}
                 H_K & -g{\Phi_0(x)}^2 \\
                 g{\Phi_0(x)}^{\ast 2} & -H_K^{\ast}
                 \end{array}
               \right) 
               \left(
                 \begin{array}{cc}
                 u_{\rm B}(x) \\ v_{\rm B}(x)
                 \end{array}
               \right)
               = \varepsilon\left(
                 \begin{array}{cc}
                   u_{\rm B}(x) \\ v_{\rm B}(x)
                 \end{array}
               \right),  \label{eq:BdGE2}\\
               H_K = -\frac{1}{2m}\frac{d^2}{dx^2}
               -{\rm i}\frac{K}{m}\frac{d}{dx}+\epsilon_K
               -\mu+V\left(x\right)+2g|\Phi_0|^2.
       \end{eqnarray}
The non-diagonal element has the periodicity of $a$ in Eq. \!(\ref{eq:BdGE2}) and the Bloch theorem can be applied to $\psi_{\rm B}(x)\equiv\left(u_{\rm B}(x),v_{\rm B}(x)\right)$ as
  \begin{eqnarray}
  \psi_{\rm B}(x+a)&=&\psi_{\rm B}(x){\rm e}^{{\rm i}qa},
  \label{eq:blochex}\\
  \left.\frac{d\psi_{\rm B}}{dx}\right|_{x+a}&=&
  \left.\frac{d\psi_{\rm B}}{dx}\right|_{x}{\rm e}^{{\rm i}qa},
  \label{eq:blochexderi}
  \end{eqnarray}
where $q$ is the quasimomentum of the excitation.
Solving Eqs. (\ref{eq:blochex}) and (\ref{eq:blochexderi}) at $x=-\frac{a}{2}$, one obtains
  \begin{eqnarray}
  {\rm exp}\left[{\rm i}\frac{(-k_1+k_2)a}{2}\right]
  +(t^{\rm l}t^{\rm r}-r^{\rm l}r^{\rm r})
  {\rm exp}\left[{\rm i}\frac{(k_1-k_2)a}{2}\right]=
  \nonumber\\
  t^{\rm l}{\rm exp}\left[{\rm i}
  \frac{(-2q+k_1+k_2)a}{2}\right]
  +t^{\rm r}{\rm exp}\left[{\rm i}
  \frac{(2q-k_1-k_2)a}{2}\right].
  \label{eq:rel_eq}
  \end{eqnarray}
The Bogoliubov band $\varepsilon(q)$ can be calculated by solving Eq. (\ref{eq:rel_eq}). Since we are assuming $a\gg \xi_{\rm av}$, the localized components around the potential barriers, which decay exponentially and vanish at $|x|=\frac{a}{2}$, do not appear explicitly in Eq. \!(\ref{eq:rel_eq}).

If $\varepsilon$ is real, one can prove from the Bogoliubov equations that the Wronskian defined by
  \begin{eqnarray}
    \!W\!(\psi^{j\ast}\!, \!\psi^i)\!=\!
    u^{j\ast}\frac{d}{dx}u^{i}\!-\!u^i\frac{d}{dx}u^{j\ast}\!
    +\!v^{j\ast}\frac{d}{dx}v^i\!-\!v^i\frac{d}{dx}v^{j\ast}
    \label{eq:wron}
  \end{eqnarray}
 is independent of $x$, where $\psi^{j}$ and $\psi^{i}$ are the solutions with the same energy.
By evaluating $W(\psi^{{\rm l}\ast}, \psi^{\rm l})$ and $W(\psi^{{\rm r}\ast}, \psi^{\rm r})$, one obtains
  \begin{eqnarray}
  \frac{-k_2(|u_{k_2}|^2+|v_{k_2}|^2)-Q}
  {k_1(|u_{k_1}|^2+|v_{k_1}|^2)+Q}
  |r^{\rm l}|^2+|t^{\rm l}|^2&=&1,
  \label{eq:tr_left}\\
  \frac{k_1(|u_{k_1}|^2+|v_{k_1}|^2)+Q}
  {-k_2(|u_{k_2}|^2+|v_{k_2}|^2)-Q}
  |r^{\rm r}|^2+|t^{\rm r}|^2&=&1,
  \label{eq:tr_right}
  \end{eqnarray}
which express the conservation law of energy flux~\cite{rf:antun}.
By evaluating $W(\psi^{{\rm r}\ast}, \psi^{\rm l})$, one also obtains
  \begin{eqnarray}
  \left(k_2\left(|u_{k_2}|^2+|v_{k_2}|^2\right)
  +Q\right)r^{\rm l}t^{{\rm r}\ast}=
  \left(k_1\left(|u_{k_1}|^2+|v_{k_1}|^2\right)
  +Q\right)r^{{\rm r}\ast}t^{\rm l}.
  \label{eq:tr_lr}
  \end{eqnarray}
Substituting Eqs. \!(\ref{eq:tr_left}), (\ref{eq:tr_right}) and (\ref{eq:tr_lr}) into Eq. \!(\ref{eq:rel_eq}), the relation between $\varepsilon$ and $q$ can be simplified as
  \begin{eqnarray}
  \frac{ {\rm cos}\left[\frac{(k_1-k_2)a}{2}
  +\frac{\delta^{\rm l}+\delta^{\rm r}}{2}\right] }{|t|}=
  {\rm cos}\left[\frac{(2q-k_1-k_2)a}{2}
  +\frac{-\delta^{\rm l}+\delta^{\rm r}}{2}\right],
  \label{eq:simprel_eq}
  \end{eqnarray}
where
  \begin{eqnarray}
  t^{\rm l}=|t|{\rm e}^{{\rm i}\delta^{\rm l}},\,\,
  t^{\rm r}=|t|{\rm e}^{{\rm i}\delta^{\rm r}}.
  \end{eqnarray}
It should be recalled that Eqs. \!(\ref{eq:rel_eq}) and (\ref{eq:simprel_eq}) are valid only when $a \gg\xi_{\rm av}$.
Equation \!(\ref{eq:simprel_eq}) is not valid when $\varepsilon$ is complex, because $\varepsilon$ is assumed to be real in the derivation. We have to solve Eq. \!(\ref{eq:rel_eq}) when complex $\varepsilon(q)$ exists.

\begin{figure}[tb]
\includegraphics[width=8cm,height=6cm]{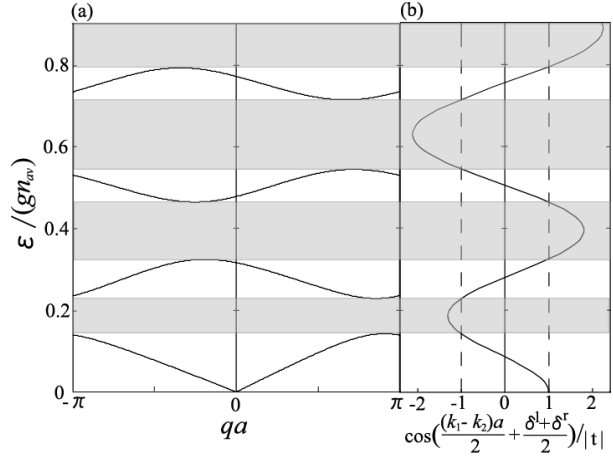}
\caption{\label{fig:exband}
Bogoliubov band of a current-carrying condensate in a KP potential for $a=15\xi_{\rm av}$, $V_0=5 gn_{\rm av}\xi_{\rm av}$ and $K=\frac{K_{\rm A}}{2}$.
The band structure is shown in (a) and the left-hand side of Eq. \!(\ref{eq:simprel_eq}) as a function of energy is shown in (b).
}
\end{figure}
Solving Eq. \!(\ref{eq:simprel_eq}), the Bogoliubov band $\varepsilon(q)$ can be calculated. The Bogoliubov band  for $a=15\,\xi_{\rm av}$, $V_0=5\,gn_{\rm av}\xi_{\rm av}$ and $K=\frac{K_{\rm A}}{2}$ is shown in Fig. \!\ref{fig:exband}(a).
Since $\varepsilon(q)$ is real and positive, the condensate with superfluid flow is stable when $K=\frac{K_{\rm A}}{2}$.
In contrast to the case of a current-free condensate in Ref. \cite{rf:wareware}, $\varepsilon(q)$ is asymmetric with respect to $q=0$. This is because the existence of the supercurrent breaks the left-right symmetry of the system.

In Fig. \!\ref{fig:exband}(b), the left-hand side of Eq. \!(\ref{eq:simprel_eq}) as a function of $\varepsilon$ is shown.
There exists no solution of Eq. \!(\ref{eq:simprel_eq}) when the absolute value of the left-hand side exceeds unity, because the absolute value of the right-hand side is always less than unity. The energy regions for no solutions correspond to the band gaps. They are expressed as the shaded regions in Fig. \!\ref{fig:exband}.

\section{Stability of superfluid flow}

\begin{figure}[tb]
      \includegraphics[width=7.0cm,height=4.5cm]{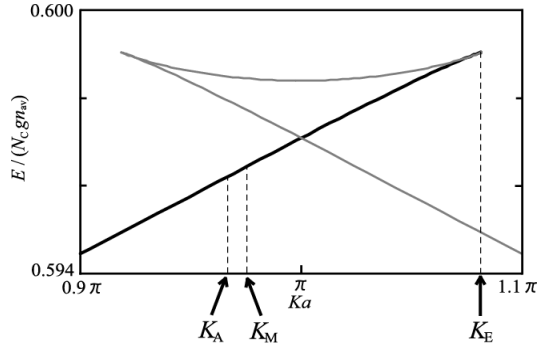}
\caption{\label{fig:mag}
Magnification of the swallow-tail energy loop in the first condensate band for $(a,V_0)=(15 \,\xi_{\rm av},5 \,gn_{\rm av}\xi_{\rm av})$ shown in Fig. \ref{fig:meanener}.
}
\end{figure}

Since the phonons around $q=0$ are expected to be closely related to the stability, we shall calculate the excitation spectrum in the limit $q\rightarrow 0$.
If one expands Eq. \!(\ref{eq:rel_eq}) assuming $q\ll \frac{\pi}{a}$, the linear dispersion of the excitation spectrum is obtained as
  \begin{eqnarray}
  \varepsilon=
  \left\{\begin{array}{cc}
  (c_{\rm b}+v_{\mu})|q|,\,\,\, q>0,\\
  (c_{\rm b}-v_{\mu})|q|,\,\,\, q<0,
  \end{array}\right.
  \label{eq:lowener_ex}
  \end{eqnarray}
where $c_{\rm b}$ and $v_{\mu}$ are given by Eqs. \!(\ref{eq:bogophonon}) and (\ref{eq:gchph}), respectively.
Equation \!(\ref{eq:lowener_ex}) has been derived previously by using the hydrodynamic analysis~\cite{rf:pesmith}, the systematic $q$-expansion of the Bogoliubov equations~\cite{rf:edo} and the tight-binding model~\cite{rf:smerzi2,rf:konabe2}.
The change in the slope of the linear dispersion by $v_{\mu}$ is analogous to the change of the phonon dispersion in a uniform system due to the Galilean transformation.


In Fig. \!\ref{fig:mag}, a magnification of a swallow-tail energy loop in the first condensate band is shown, and $K_{\rm A}$, $K_{\rm M}$ and $K_{\rm E}$ are indicated by allows. 
We first consider a condensate with a positive group velocity $v_{\rm g}$ in the lower portion of the swallow-tail (the region shown by the black solid line in Fig. \ref{fig:mag}).
The Landau (energetical) instability occurs if there exists any excitation with negative energy.
One can clearly see from Eq. \!(\ref{eq:lowener_ex}) that the Landau instability occurs if $v_{\mu}>c_{\rm s}$. We confirmed the fact that the onset of the Landau instability is due to excitations with small $q$ by numerically solving Eq. \!(\ref{eq:rel_eq}).
Therefore, the onset of the Landau instability is given by the condition
  \begin{eqnarray}
  v_{\mu}=c_{\rm s}.
  \label{eq:chev_phov}
  \end{eqnarray}
Assuming $V_0\gg gn_{\rm av}\xi_{\rm av}$ and solving Eq. \!(\ref{eq:chev_phov}), one obtains
  \begin{eqnarray}
  {\rm cos}\varphi_{\rm Lan}\simeq \frac{gn_{\rm av}\xi_{\rm av}}{V_0}
  +\frac{gn_{\rm av}\xi_{\rm av}^2}{aV_0}
  -\frac{(gn_{\rm av}\xi_{\rm av})^2}{2V_0^2},
  \label{eq:philan}
  \end{eqnarray}
where $\varphi_{\rm Lan}$ is the phase difference corresponding to the critical quasimomentum for the Landau instability $K_{\rm Lan}$.
From Eqs. \!(\ref{eq:phiano}) and (\ref{eq:philan}), one sees that $K_{\rm Lan}=K_{\rm A}$, namely the onset of the Landau instability coincides with the point at which the anomalous tunneling is absent.

Dynamical instability occurs if there exists any excitation with complex energy.
From Eqs. \!(\ref{eq:bogophonon}) and (\ref{eq:lowener_ex}), one can see that excitations around $q=0$ cause the dynamical instability if the effective mass is negative, namely $K>K_{\rm M}$.
On the other hand, it can be expected that the dynamical instability sets in at $K<K_{\rm M}$ due to excitations around $q=\frac{\pi}{a}$, because crossing of phonon and anti-phonon branches, which is crucial to the appearance of excitations with complex energies, occurs initially at $q=\frac{\pi}{a}$~\cite{rf:wuniu1,rf:wuniu2,rf:edo}.
We calculate the critical value of $K$ for the dynamical instability caused by excitations around $q=\frac{\pi}{a}$.
The first Bogoliubov band is analytically expressed by assuming $K\sim K_{\rm A}$.
For avoiding the complication of equations, we introduce a variable $b$ defined as
  \begin{eqnarray}
  b=\frac{aV_0}{gn_{\rm av}\xi_{\rm av}}\left(
  {\rm cos}\varphi_{\rm A}-{\rm cos}\varphi\right),
  \label{eq:bee}
  \end{eqnarray}
where $-2\leq b\leq 2$.
$b$ is a function of $K$, for example, $b(K_{\rm A})=0$ and $b(K_{\rm M})=2$.
Assuming $V_0\gg gn_{\rm av}\xi_{\rm av}$ and substituting Eq. \!(\ref{eq:bee}) into Eq. \!(\ref{eq:rel_eq}), one obtains
  \begin{eqnarray}
  \frac{\varepsilon}{gn_{\rm av}}
  \simeq\frac{gn_{\rm av}\xi_{\rm av}^2}{aV_0}\left({\rm sin}(qa)
  \pm\sqrt{{\rm sin}^2(qa)
  -2b\,{\rm sin}^2\left(\frac{qa}{2}\right)}\right).
  \label{eq:bet_landy}
  \end{eqnarray}
Expanding Eq. \!(\ref{eq:bet_landy}) around $q=\frac{\pi}{a}$, it reduces to
  \begin{eqnarray}
  \frac{\varepsilon}{gn_{\rm av}}\simeq 
  \frac{gn_{\rm av}\xi_{\rm av}^2}{a V_0}\left(\pi-qa+\sqrt{-2b}\right).
  \label{eq:around_ka}
  \end{eqnarray}
One can see from Eq. \!(\ref{eq:around_ka}) that the dynamical instability due to excitations with $q=\frac{\pi}{a}$ starts at $K=K_{\rm A}$.
Therefore, both the Landau and dynamical instabilities start to occur at $K=K_{\rm A}$.
In the case of condensates in a sinusoidal potential, the onset of the Landau instability asymptotically approaches that of the dynamical instability in the limit of $gn_{\rm av}\gg E_{\rm R}$, where $E_{\rm R}=\frac{\pi^2}{2m a^2}$ is the recoil energy~\cite{rf:pesmith}.
Since our assumption of sufficiently large lattice constant compared to the healing length corresponds to the condition $gn_{\rm av}\gg E_{\rm R}$, our result is consistent with that of condensates in a sinusoidal potential.

\begin{figure}[tb]
      \includegraphics[width=7.0cm,height=4.5cm]{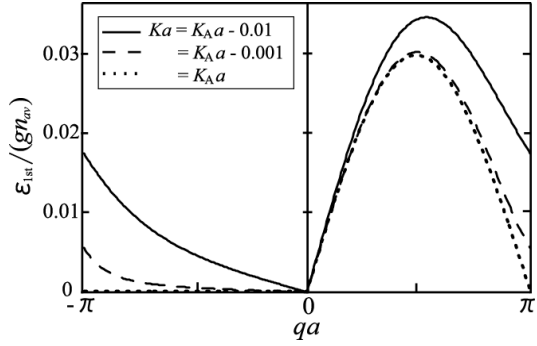}
\caption{\label{fig:bLan}
First Bogoliubov band with $K\le K_{\rm A}$ for $(a,V_0)=(15 \,\xi_{\rm av},5 \,gn_{\rm av}\xi_{\rm av})$.
}
\end{figure}
\begin{figure}[tb]
      \includegraphics[width=7.0cm,height=9cm]{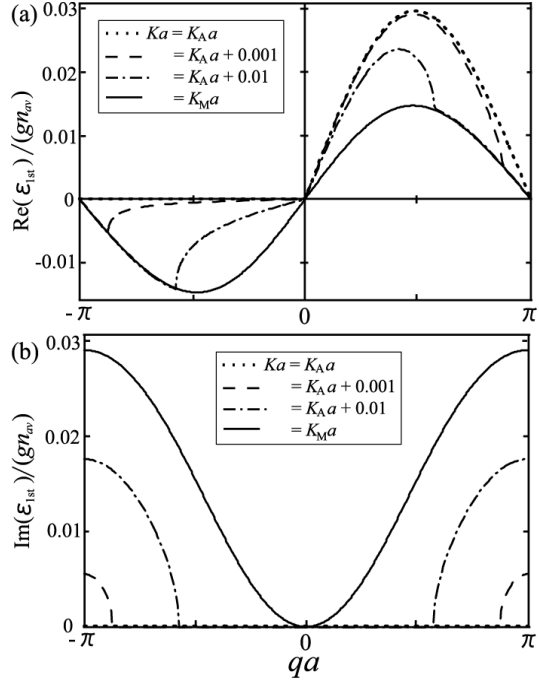}
\caption{\label{fig:aLan}
First Bogoliubov band with $K_{\rm A}\le K\le K_{\rm M}$ for $(a,V_0)=(15\,\xi_{\rm av},5 gn_{\rm av}\xi_{\rm av})$.
(a) The real part of the first Bogoliubov band.
(b) The imaginary part of the first Bogoliubov band.
}
\end{figure}
In Figs. \!\ref{fig:bLan} and \ref{fig:aLan}, the first Bogoliubov bands in the regions $K\le K_{\rm A}$ and $K_{\rm A}\le K\le K_{\rm M}$ are shown, respectively.
In these figures, one can clearly see that the Landau and dynamical instabilities start simultaneously at $K=K_{\rm A}$.
It is also clear that condensates in the region of $K_{\rm A}<K<K_{\rm M}$ are always unstable due to both Landau and dynamical instabilities.
As $K$ increases and approaches $K_{\rm A}$, the slope of the linear dispersion reduces. The slope becomes zero at $K=K_{\rm A}$ (Fig. \ref{fig:bLan}).
When $K$ exceeds $K_{\rm A}$, excitations with negative energies appear (Fig. \ref{fig:aLan}(a)).
At the same time, the imaginary part of the excitation energy grows around $q=\frac{\pi}{a}$ (Fig. \ref{fig:aLan}(b)).
For $K>K_{\rm M}$, all the excitations in the first Bogoliubov band have complex energies.

\begin{figure}[tb]
      \includegraphics[width=7.0cm,height=9.0cm]{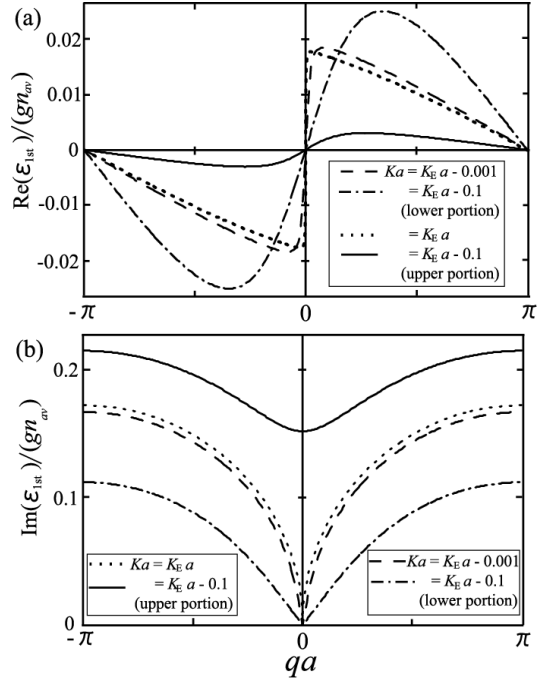}
\caption{\label{fig:edge1st}
First Bogoliubov band around the edge of the swallow-tail for $(a,V_0)=(15,5)$.
(a) The real part of the first Bogoliubov band.
(b) The imaginary part of the first Bogoliubov band.
}
\end{figure}
\begin{figure}[tb]
      \includegraphics[width=7.0cm,height=4.5cm]{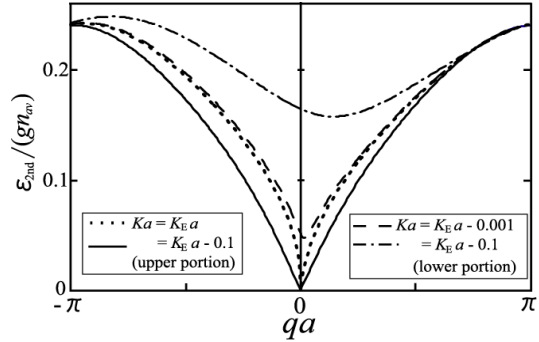}
\caption{\label{fig:edge2nd}
Second Bogoliubov band around the edge of the swallow-tail for $(a,V_0)=(15,5)$.
}
\end{figure}
We consider a condensate with a positive group velocity $v_{\rm g}$ around the edge of the swallow-tail $K=K_{\rm E}$, including the upper portion of it.
In Figs. \!\ref{fig:edge1st} and \ref{fig:edge2nd}, the first and second Bogoliubov bands around $K=K_{\rm E}$ are shown, respectively.
Since there always exist excitations with complex energies in the first Bogoliubov band (Fig. \ref{fig:edge1st}), the upper portion of the swallow-tail is dynamically unstable.
This result agrees with that of the numerical simulation by the time-dependent GP equation in Ref.~\cite{rf:seaman1}.
On the other hand, we find in Fig. \ref{fig:edge2nd} that there also exists a phonon spectrum in the upper portion, reflecting the positive effective mass.
As $K$ in the lower portion of the swallow-tail approaches $K_{\rm E}$, the bottom of the second Bogoliubov band approaches the origin.
The second Bogoliubov band turns into a gapless dispersion when $K$ reaches $K_{\rm E}$, and it has a phonon spectrum in the upper portion.
\begin{figure}[tb]
      \includegraphics[width=7.5cm,height=4.5cm]{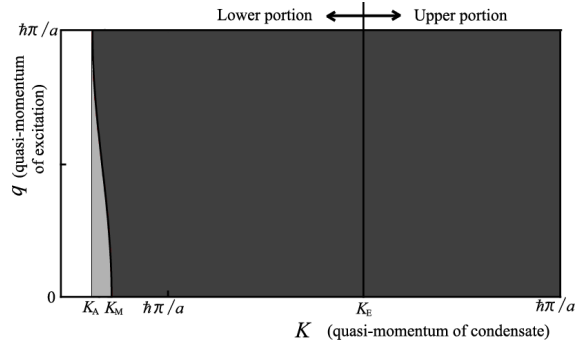}
\caption{\label{fig:stability}
Stability phase diagram of the condensate in a KP potential when $(a,V_0)=(15\,\xi_{\rm av},5\,gn_{\rm av}\xi_{\rm av})$.
The light shaded area and the dark shaded area correspond to the regions of the Landau instability and the dynamical instability, respectively.
This figure is focused to the swallow-tail energy loop.
}
\end{figure}

Thus, the stability of the condensate in a KP potential has been revealed in all the regions of the first condensate band.
To conclude this section, we show the stability phase diagram in Fig. \ref{fig:stability}.

\section{CONCLUSION}
In summary, we have studied the stability and excitations of Bose-Einstein condensates with superfluid current in a Kronig-Penney potential based on the Gross-Pitaevskii mean-field theory. 
Solving the time-independent Gross-Pitaevskii equation, analytic form of the condensate wave function in the first condensate band has been obtained.
Using this solution, the condensate band structure, group velocity and the effective mass have been calculated.
It has been found that the first condensate band has a swallow-tail energy loop if $2V_0<gn_{\rm av} a$.
This condition agrees with the result obtained by numerical calculations in Ref.~\cite{rf:seaman1}.

Imposing the Bloch theorem on the solution of the Bogoliubov equations in the single-barrier problem, we have calculated the Bogoliubov band and investigated the stability of the condensate.
We have found that the onsets of the Landau instability caused by excitations around $q=0$ and the dynamical instability caused by excitations around $q=\frac{\pi}{a}$ coincide, and anomalous tunneling of low-energy phonons for each potential barrier is absent at this point.
When the effective mass is negative, condensates are dynamically unstable due to the excitations in the first Bogoliubov band.
It has been found that the second Bogoliubov band has a phonon spectrum in the upper portion of a swallow-tail, while the upper portion is always dynamically unstable due to the excitations in the first Bogoliubov band.
\begin{acknowledgments}
The authors would like to thank S. Kurihara, N. Yokoshi, S. Konabe, and T. Nikuni for fruitful comments and discussions.
They also greatly acknowledge useful comments of D. L. Kovrizhin.
I.D. is grateful for insightful discussions with L. D. Carr and C. W. Clark.
I.D. is supported by a Grant-in-Aid from JSPS.

\end{acknowledgments}
\appendix
\section*{Appendix: Analytic expressions of physical quantities}
In this appendix, we show the higher order terms of the physical quantities expanded into the power series of $gn_{\rm av}\xi_{\rm av}/V_0$ and $\xi_{\rm av}/a$.
For simplicity, we introduce the following notation,
  \begin{equation}
  {\rm c} = {\rm cos}\varphi,\tag{A1}
  \end{equation}
  \begin{equation}
  {\rm s} = {\rm sin}\varphi,\tag{A2}
  \end{equation}
  \begin{equation}
  \tilde{V}_0 = \frac{V_0}{gn_{\rm av}\xi_{\rm av}},\tag{A3}
  \end{equation}
  \begin{equation}
  \tilde{a} = \frac{a}{\xi_{\rm av}},\tag{A4}
  \end{equation}
  \begin{equation}
  \tilde{K} = K\xi_{\rm av}.\tag{A5}
  \end{equation}
The superfluid momentum $Q$, $\gamma(ja)$, chemical potential $\mu$, energy of the condensate $E$, group velocities $v_{\rm g}$ and $v_{\mu}$, and effective mass $m^{\ast}$ are expressed as
  \begin{equation}
  \tilde{Q}=Q \xi_{\rm av}\simeq \frac{\rm s}{2\tilde{V}_0}
  \left(1+\frac{1}{\tilde{a}}+\frac{\rm c}{\tilde{V}_0}+\frac{2}{\tilde{a}^2}
  +\frac{-1+2{\rm c}}{\tilde{a}\tilde{V}_0}
  +\frac{-2-{\rm c}+3{\rm c}^2}{2\tilde{V}_0^2}\right),
  \tag{A6}\label{eq:Qho}
  \end{equation}
  \begin{equation}
  \gamma(ja) \simeq \frac{1+{\rm c}}{2\tilde{V}_0}
  \left(1+\frac{1}{\tilde{a}}+\frac{-1-{\rm c}}{\tilde{V}_0}
  +\frac{1}{2\tilde{a}^2}+\frac{-5+3{\rm c}}{2\tilde{a}\tilde{V}_0}
  +\frac{-1-4{\rm c}+3{\rm c}^2}{2\tilde{V}_0^2}\right),
  \tag{A7}
  \end{equation}
  \begin{equation}
  \tilde{\mu}=\frac{\mu}{gn_{\rm av}}\simeq \tilde{\mu}|_{\varphi=0}
  +(1-{\rm c})\left(\frac{1}{\tilde{a}\tilde{V}_0}
  +\frac{1+{\rm c}}{8\tilde{V}_0^2}+\frac{3}{\tilde{a}^2\tilde{V}_0}
  +\frac{5+5{\rm c}}{4\tilde{a}\tilde{V}_0^2}+\frac{\rm c+c^3}{4\tilde{V}_0^3}
  \right. \nonumber
  \end{equation}
  \begin{equation}
  \left.
  +\frac{4}{\tilde{a}^3\tilde{V}_0}+\frac{1+3{\rm c}}{\tilde{a}^2\tilde{V}_0^2}
  +\frac{-7+5{\rm c}+8{\rm c}^2}{4\tilde{a}\tilde{V}_0^3}
  +\frac{-2-3{\rm c}+3{\rm c}^2+4{\rm c}^3}{8\tilde{V}_0^4}\right),
  \tag{A8}
  \end{equation}
  \begin{equation}
  \tilde{\mu}|_{\varphi=0}\simeq 1+\frac{2}{\tilde{a}}+\frac{2}{\tilde{a}^2}
  -\frac{2}{\tilde{a}\tilde{V}_0}+\frac{1}{\tilde{a}^3}
  -\frac{6}{\tilde{a}^2\tilde{V}_0}-\frac{8}{\tilde{a}^3\tilde{V}_0}
  +\frac{4}{\tilde{a}^2\tilde{V}_0^2}+\frac{2}{\tilde{a}\tilde{V}_0^3},
  \tag{A9}
  \end{equation}
  \begin{equation}
  \tilde{E}=\frac{E}{N_{\rm C}gn_{\rm av}}\simeq\tilde{E}|_{\varphi=0}
  +(1-{\rm c})\left(\frac{1}{2\tilde{a}\tilde{V}_0}
  +\frac{1+{\rm c}}{8\tilde{V}_0^2}+\frac{2}{\tilde{a}^2\tilde{V}_0}
  +\frac{1+{\rm c}}{\tilde{a}\tilde{V}_0^2}+\frac{\rm c+c^3}{4\tilde{V}_0^3}
  \right. \nonumber
  \end{equation}
  \begin{equation}
  \left.
  +\frac{4}{\tilde{a}^3\tilde{V}_0}+\frac{2+3{\rm c}}{\tilde{a}^2\tilde{V}_0^2}
  +\frac{-5+7{\rm c}+10{\rm c}^2}{6\tilde{a}\tilde{V}_0^3}
  +\frac{-2-3{\rm c}+3{\rm c}^2+4{\rm c}^3}{8\tilde{V}_0^4}\right),
  \tag{A10}
  \end{equation}
  \begin{equation}
  \tilde{E}|_{\varphi=0}\simeq \frac{1}{2}+\frac{4}{3\tilde{a}}
  +\frac{2}{\tilde{a}^2}-\frac{1}{\tilde{a}\tilde{V}_0}+\frac{2}{\tilde{a}^3}
  -\frac{4}{\tilde{a}^2\tilde{V}_0}+\frac{4}{3\tilde{a}^4}
  -\frac{8}{\tilde{a}^3\tilde{V}_0}
  +\frac{2}{\tilde{a}^2\tilde{V}_0^2}+\frac{2}{3\tilde{a}\tilde{V}_0^3}.
  \tag{A11}
  \end{equation}
  \begin{equation}
  \frac{v_{\rm g}}{c_{\rm s}}=
  \frac{\partial\tilde{E}}{\partial\tilde{K}}\simeq 
  \frac{\rm s}{2\tilde{V}_0}
  \left(1+\frac{4}{\tilde{a}}+\frac{\rm c}{\tilde{V}_0}+\frac{8}{\tilde{a}^2}
  +\frac{\rm -2+3c}{\tilde{a}\tilde{V}_0}+\frac{\rm -2-c+3c^2}{2\tilde{V}_0^2}
  \right),
  \tag{A12}\label{eq:vgho}
  \end{equation}
  \begin{equation}
  \frac{v_{\mu}}{c_{\rm s}}=
  \frac{\partial\tilde{\mu}}{\partial\tilde{K}}\simeq
  \frac{\rm s}{\tilde{V}_0}\left(\frac{1}{\tilde{a}}
  +\frac{\rm c}{4\tilde{V}_0}+\frac{3}{\tilde{a}^2}
  +\frac{\rm 5c}{2\tilde{a}\tilde{V}_0}
  +\frac{\rm -1+2c-3c^2+4c^3}{4\tilde{V}_0^2}\right.
  +\frac{4}{\tilde{a}^3}\nonumber
  \end{equation}
  \begin{equation}
  \left.+\frac{\rm -2+6c}{\tilde{a}^2\tilde{V}_0}
  +\frac{\rm -6-3c+12c^2}{2\tilde{a}\tilde{V}_0^2}
  +\frac{\rm 1-12c-3c^2+16c^3}{8\tilde{V}_0^3}
  \right)\bigg/\frac{\partial\tilde{K}}{\partial\varphi},
  \tag{A13}
  \end{equation}
  \begin{equation}
  \frac{m^{\ast}}{m}=
  \left(\frac{\partial^2 \tilde{E}}{\partial \tilde{K}^2}\right)^{-1}\simeq
  \frac{\frac{\partial\tilde{K}}{\partial\varphi}}
  {\frac{\rm c}{2\tilde{V}_0}+\frac{\rm 2c}{\tilde{a}\tilde{V_0}}
  +\frac{\rm -1+2c^2}{\tilde{V}_0^2}+\frac{\rm 4c}{\tilde{a}^2\tilde{V}_0}
  +\frac{\rm -3-2c+6c^2}{2\tilde{a}\tilde{V}_0}
  +\frac{1-8c-2c^2+9c^3}{4\tilde{V}_0^3}
  },
  \tag{A14}\label{eq:effmho}
  \end{equation}
where
  \begin{equation}
  \frac{\partial\tilde{K}}{\partial\varphi}\simeq\frac{1}{\tilde{a}}
  +\frac{\rm c}{2\tilde{V}_0}+\frac{\rm c}{\tilde{a}\tilde{V}_0}
  +\frac{\rm -1+2c^2}{2\tilde{V}_0^2}+\frac{c}{\tilde{a}^2\tilde{V}_0}
  +\frac{\rm -2-c+4c^2}{2\tilde{a}\tilde{V}_0^2}
  +\frac{\rm 1-8c-2c^2+9c^3}{4\tilde{V}_0^3}.
  \tag{A15}
  \end{equation}
Equations (\ref{eq:Qho}), (\ref{eq:vgho}) and (\ref{eq:effmho}) are necessary for the calculation of Eqs. \!(\ref{eq:phiano}), (\ref{eq:phimax}) and (\ref{eq:phiE}), respectively.


\begin{thebibliography}{99}
\bibitem{rf:lastnote}
The quasimomentum $K_{\rm A}$, which gives the maximum value of the superfluid momentum $Q$, does not coincide with the quasimomentum $K_{\rm M}$, which gives the maximum value of the supercurrent $\frac{n_0 Q}{m}$, because the center density $n_0$ of the condensate also depends on the quasimomentum $K$.
\end{thebibliography}
\end{document}